\documentclass{aa}  

\usepackage{color}
\usepackage{graphicx}
\usepackage{array}
\usepackage{txfonts}
\AddToHook{begindocument/before}{\usepackage{hyperref}}
\begin{document}

   \title{DULAG: A DUal and Lensed AGN candidate catalog with GMP method}


   \author{Qiqi Wu
          \inst{1,2}
          \and
          M. Scialpi\inst{3,4,5}
          \and
          Shilong Liao\inst{1,2} \thanks{Corresponding author,  \email{shilongliao@shao.ac.cn}}
          \and
          F. Mannucci\inst{5} \thanks{Corresponding author,  \email{filippo.mannucci@inaf.it}}
          \and
          Zhaoxiang Qi\inst{1,2} \thanks{Corresponding author,  \email{zxqi@shao.ac.cn}}
          }

   \institute{Shanghai Astronomical Observatory, Chinese Academy of Sciences, Shanghai 200030, China.
         \and{University of Chinese Academy of Sciences, Beijing 100049, China.}
        \and{University of Trento, Via Sommarive 14, I-38123 Trento, Italy.  }
        \and{Università di Firenze, Dipartimento di Fisica e Astronomia, via G. Sansone 1, 50019 Sesto F.no, Firenze, Italy. }
        \and{}
        INAF - Osservatorio Astrofisico di Arcetri, largo E. Fermi 5, 50125 Firenze, Italy.\\
        }

   \date{Received XXXX, XXXX; accepted XXXX, XXXX}

 
  \abstract
   {A series of studies have demonstrated that the 
    $Gaia$ multipeak method (GMP) is a very efficient technique
    to select active galactic nucleus (AGN) pair candidates. 
    The number of candidates is determined by the size of the input AGN catalogs, usually limited to spectroscopically-confirmed objects.
    }
   {The objective of this work is to compile a larger and highly reliable catalog of GMP pair candidates extracted from the six million objects the $Gaia$ AGN catalog, the majority of which lack spectroscopic information.}
   {In order to ascertain the differences in the properties of GMP pair candidates compared to normal AGN, we conducted an investigation utilising samples of GMP AGN. These differences were employed to establish the optimal selecting criteria, which ultimately led to the identification of a highly reliable candidate catalog.}
   {We found significant differences in astrometry and multi-band colour distribution between normal AGN and GMP pair candidates. A DUal and Lensed AGN candidate catalog with GMP method (DULAG) comprising 5,286 sources was ultimately compiled, accompanied by a highly reliable Golden sample of 1,867 sources. A total of 37 sources in the Golden sample have been identified as dual AGN or lensed AGN. For the majority of sources in the Golden sample, we provide reference redshifts and find three close AGN pair candidates among them.}
   {}


   \maketitle
%

\section{Introduction}
\label{introduction}
Supermassive black hole binaries (SMBHBs) are the key to our understanding of the evolutionary process of the massive objects in the Universe \citep[e.g.][]{dosopoulou2017dynamical}, the verification of gravitational theories \citep[e.g.][]{healy2012late}, and the origin of the gravitational waves \citep[e.g.][]{belczynski2014formation}. \citet{rodriguez2006compact} discovered a SMBHB with a projected separation of 7.3 pc using the Very Long Baseline Array (VLBA), which is the closest black hole pair yet. Many SMBHB candidates have been reported over the years \citep[e.g.][]{ju2013search,graham2015systematic,graham2015possible,wang2017searching, liu2019supermassive}. However, it's difficult to identify them because the components of the systems are too close to be distinguished by the telescope. 

Dual AGNs at kpc separations are the precursor of SMBHBs, with more easily observable separations, so it is  important to take systematic investigations and observations of them. In the past decade, a large number of AGN pairs (dual and lensed AGNs) with different separations have been discovered. In particular, with the release of high-resolution data from $Gaia$ DR2 \citep{brown2018gaia} and (E)DR3 \citep{brown2021gaia,vallenari2023gaia}, many authors have developed effective and systematic methods to search for AGN pairs using $Gaia$ data. \citet{shen2019varstrometry} proposed that AGN pair candidates can be selected by varstrometry method, which is a strong testament to the effectiveness of $Gaia$'s data in the search for Dual AGNs \citep[e.g.][]{chen2022varstrometry,li2023varstrometry}.

Recently, \citet{mannucci2022unveiling} proposed the Gaia Multi Peak (GMP) method for effectively selecting dual AGN candidates. The GMP method select AGN pair candidates by using the $Gaia$ multipeak parameter $ipd\_frac\_multi\_peak$ (FMP). The basic principle is that when $Gaia$ is observing pairs with angular separations between $~0.15^{\prime \prime}$  and $~0.8^{\prime \prime}$, if the telescope cannot resolve them effectively, a secondary peak might be detected in the profile. The FMP parameter reports the  percentage (from 0 to 100) of $Gaia$ scans showing the presence of multiple peaks.
\citet{mannucci2023gmp} have shown that
for sources with $G \textless 20.5$ mag and $\delta$ (separation) $\textgreater 0.15^{\prime \prime}$, the cut of FMP $\geq 8$ ensures a low contamination and high efficiency. Based on this method and high-precision spectroscopic observations, many AGN pairs have been identified \citep[e.g.][]{ciurlo2023new,scialpi2023muse}. 

The purity of the AGN pair candidates selected by, varstrometry, GMP, or other astrometric methods \citep[e.g.][]{wu2022squab}, can be improved by considering the quasars that have already been spectroscopically identified by different telescopes, such as the Sloan Digital Sky Survey quasar catalog \citep[SDSS,][]{lyke2020sloan}, the LAMOST quasar survey \citep{dong2018large}, the DESI spectroscopic survey \citep{chaussidon2023target}, and so on. The Million Quasars (Milliquas) catalog \citep{flesch2023million}, collects almost all the quasars currently identified for large-sample spectroscopy have been cataloged, with a total number of 0.9 million. The benefits of using these catalogs as input catalogs are obvious, as all quasars are spectroscopically identified. However, these catalogs are incomplete for most of the sky region \citep[see][for the sky coverage of these catalogs]{flesch2023million}. Therefore in order to expand the sample of AGN pair candidates, it is necessary to select reliable quasar candidates from the sources without spectroscopic observations.

Over the past decade, many outstanding projects aimed at selecting quasar candidates have been completed \citep[e.g.][]{secrest2015identification, bailer2019quasar}. Various groups have obtained large numbers of quasar candidates based on different methods, bringing the number of quasar candidates into several millions. Especially after the publication of the $Gaia$ DR3, with its high-precision astrometric data with multi-color magnitudes, \citet{gaia2023gaia} obtained a high-completeness but low-purity sample of quasar candidates numbering up to 6.6 million by multiple methods. Based on this catalog, some pure quasar candidate catalogs have been constructed. For example, \citet{storey2023quaia} built a pure QUAIA catalog of 1.3 million quasar candidates based on the $Gaia$ proper motions and $unWISE$ color \citep{lang2014unwise}, while \citet{fu2023catnorth} constructed the CatNorth catalog of 1.5 million quasar candidates ($\textgreater90\%$ purity) with data from $Gaia$, Pan-STARRS1 \citep{flewelling2020pan}, and CatWISE2020 \citep{marocco2021catwise2020}. An important objective in the creation of these substantial collections of QSO candidates was the establishment of Gaia optical celestial reference frame \citep{klioner2022gaia}. Therefore, principal characteristics of these catalogs are their high purity and the almost zero parallax and proper motion of their sources. When selecting AGN pair candidates using these catalogs, the high purity may result in a significant loss of completeness. The reason is that AGN pairs are different from normal quasars in various ways:
\begin{enumerate}
    \item Proper motions and parallaxes. It is generally accepted that, since quasars are extremely distant, their parallaxes and proper motions should be consistent with 0. Therefore, most quasar candidates selection techniques set up cut-off parallaxes and proper motions \citep[e.g.][]{lindegren2018gaia,klioner2022gaia,wu2023catalog}. However, AGN pairs may show observable parallaxes and proper motions due to apparent movements of their optical centers caused by the variability of the two AGNs \citep{makarov2022quasars}.
    \item The optical color and infrared color. $Gaia$ photometry is optimized for point sources, and this can lead to shifts in the photometry of some dual and lensed AGNs. Also, since the blue and red photometry is derived from extended spectra \citep{de2023gaia}, the $bp-g$ and $g-rp$, which are commonly used in the selection of quasar candidates, may not be reliable for sources with components. As for the infrared color, there are no known differences that exist between AGN pairs and normal AGNs when z $\textgreater 0.5$, so we will explore and discuss that in this paper.
\end{enumerate}

This paper is organized as follows. In Section \ref{s_data}, we introduce the data we used and analyze their characters. In Section \ref{s_method}, we explore the differences in astrometric and colors between GMP AGN pair candidates and normal AGNs and determine the final selection criteria. We also describe in detail the construction of our dual and lensed AGN candidate catalog (DULAG) in this section. The properties of the catalog are demonstrated in Section \ref{s_character}. In Section \ref{s_conclusion}, we give the conclusion and discuss the future of the selection.


\section{Data}
\label{s_data}
In this section we present our input catalog for AGN pair selection, which is composed of the $Gaia$ quasar candidate catalog and the Milliquas catalog. We also introduce an AGN pair candidate catalog confirmed by observation, which provides an important base for our selection.

\subsection{The $Gaia$ quasar candidate catalog}
As mentioned above, there has been a lot of work done on quasar selection. However, many of them always use strict astrometric, infrared or optical color criteria to improve the purity of the samples. $Gaia$ has released a catalog of 6.6 million quasar candidates (the \texttt{qso\_candidates} table) \citep{gaia2023gaia}, which is an exception. This catalog is composed by multiple samples that are selected based on $Gaia$ spectrum, colors, astrometric parameters, photometric light curves, surface brightness and so on. In the process of generating this catalog, the authors aim at providing a sample that is as complete as possible. As a major contributor to the catalog, the Discrete Source Classifier (DSC) module \citep{delchambre2023gaia} provides about 5.5 million sources. The DSC module contains three classifiers: Specmod, Allosmod and Combmod. Specmod and Allosmod provide the classification probabilities of the sources using BP/RP spectra and other data of the objects, respectively. Combmod gives the classification probabilities by combining the above two classifiers. The \href{https://gea.esac.esa.int/archive/documentation/GDR3/Data_analysis/chap_cu8par/sec_cu8par_apsis/ssec_cu8par_apsis_dsc.html}{online documentation} shows that when sources with quasar probability of Combmod greater than 0.5 are selected, about 5.2 million sources are available, corresponding to a completeness of about 90\%. In addition to these sources, the \texttt{qso\_candidates} table contains an additional 0.3 million sources filtered by Specmod and Allosmod, and 1.1 million sources filtered by other modules (e.g., Variability module, $Gaia$-CRF3 module, etc, all modules with purity higher than 90\%). Therefore, using this table as an major input catalog when performing AGN pair candidates selection could ensure a very high degree of completeness.

\begin{figure}
  \resizebox{\hsize}{!}{\includegraphics{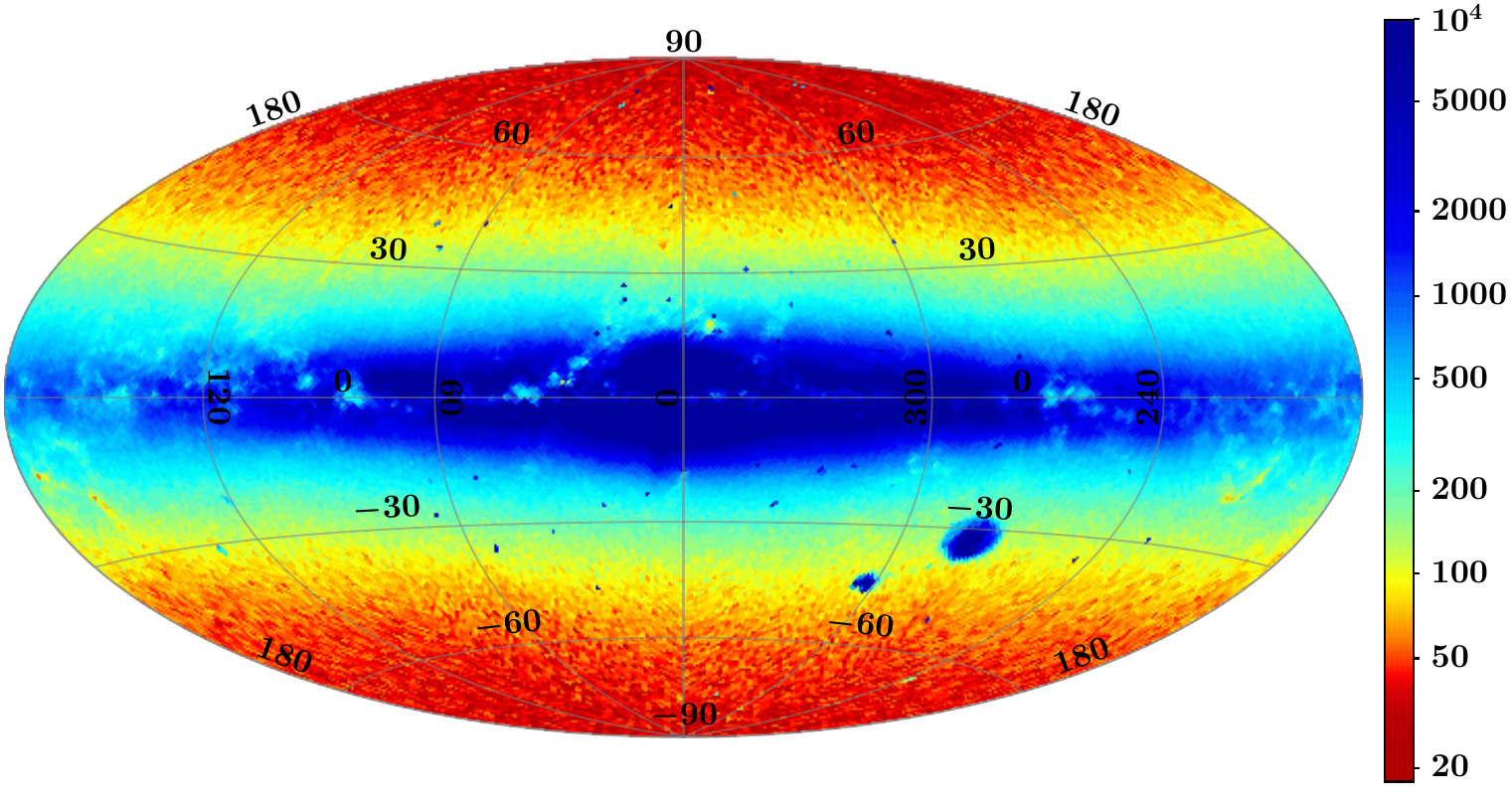}}
  \caption{The density distribution of $Gaia$ sources with FMP $\geq 8$ and $G \textless 20.5$ mag, using the Hammer Aitoff projection in Galactic coordinate. The cell of this map is approximately 0.84 deg$^2$, and the color shader shows the number of the sources in each cell.}
  \label{gmp8g205}
\end{figure}

\begin{figure}
  \resizebox{\hsize}{!}{\includegraphics{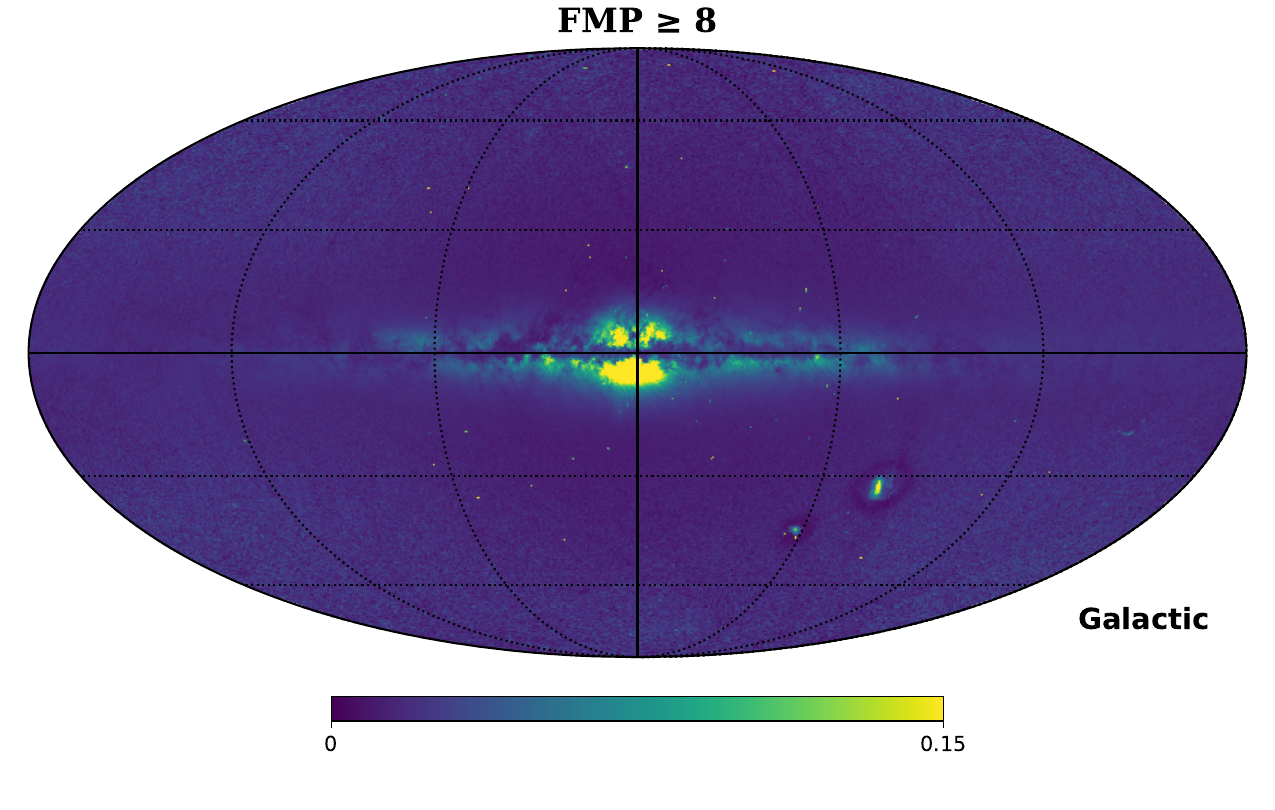}}
  \caption[Caption for LOF]{Probability map at HEALPix level 7\protect\footnotemark for the same sources as in Fig. \ref{gmp8g205}. The Probability is calculated by $k/n \times 100\%$, where k is the number of sources meeting the FMP cut ($\geq 8$) and G cut ($\textless 20.5$) in each pixel, n is the number of sources with $G \textless 20.5$ in each pixel.}
  \label{gmp8_prob}
\end{figure}

The \texttt{qso\_candidates} table covers the whole sky, including crowded regions such as the Galactic Plane and Large and Small Magellanic Clouds (LMC and SMC). However, crowded foreground stars are very serious contaminants when selecting GMP AGN pair candidates. The reasons are described below.
\begin{enumerate}
      \item Excessive stellar density in these regions can lead to serious errors when matching with other catalogs \citep{klioner2022gaia}, and matching errors will reduce the reliability of selection procedure.
      \item These regions will suffer more severe interstellar extinction than others, which will make the color of the extragalactic sources very different from other regions (high Galactic latitude) \citep{fu2021finding}.
      \item For the AGN pairs selected by GMP method, the most important parameter is FMP, which, as we described earlier, characterises the probability that a secondary peak is detected in the source light profile. In these crowded regions, the angular separation between objects is very small, resulting in a large percentage of sources having large FMP. This can lead to contamination of the GMP AGN pair samples by stellar pairs or star-quasar pairs, which we want to avoid. Fig. \ref{gmp8g205} shows the sky density of sources meeting the GMP method cut, the total number is 83,116,497, and the LMC, SMC and Galactic Plane show significantly high density. We also estimate the probability of these sources using the method named \texttt{SubsampleSelectionFunction} presented by \citet{castro2023estimating}. \texttt{SubsampleSelectionFunction} could calculate the posterior probability of a source satisfying the user-defined criteria. We use this method to calculate the probability of sources that have FMP $\geq 8$ in the sample of sources brighter than 20.5 mag, see Fig. \ref{gmp8_prob}. The result also shows that sources in these crowded regions are more likely to have high FMP values.    
   \end{enumerate}
Therefore, we remove these regions as follows: we remove circular regions of radius of $9^{\circ}$ and $6^{\circ}$ around the LMC and SMC, respetively, using the method recommended in \citet{gaia2023gaia}, which is \texttt{WHERE 1!=CONTAINS(POINT(’ICRS’, 81.3, -68.7),CIRCLE(’ICRS’, ra, dec, 9)) AND 1!=CONTAINS(POINT(’ICRS’, 16.0, -72.8), CIRCLE(’ICRS’, ra, dec, 6))}. We also exclude the regions around the Galactic Plane by set $\lvert b \rvert \textgreater 11.54^{\circ}$, the same as in \citet{delchambre2023gaia}. After this, we obtain 3,479,889 sources as our major input catalog.

\footnotetext{HEALPix is a method of subdividing the sky into pixels. In this figure, we divided the sky to 196,608 cells by level 7. Details to see the \href{https://healpix.sourceforge.io/}{online document}.{\label{pixel}}}

\subsection{The Milliquas catalog}
The Milliquas catalog is an ambitious project that aims to collect all the quasars that have been identified in the literature and is constantly updated with the latest publications. The project started in 2009 and released its final version (version 8) in 2023 \citep{flesch2023million}. The \href{https://heasarc.gsfc.nasa.gov/W3Browse/all/milliquas.html}{NASA HEASARC} provides a detailed version log for this catalog. 

The Milliquas version 8 (v8 hereafter) includes almost all quasars published as of June 2023, as well as a number of high-confidence quasar candidates, and is therefore an ideal input catalog for selecting GMP AGN pairs. Table \ref{milliquas_n} displays the variation in the number of sources between different versions of the Milliquas catalog  (only some of the versions with detailed publications are shown). Both version 5.2 \citep[v5.2 ][]{2017yCat.7280....0F} and version 6.3 \citep[v6.3 ][]{2019yCat.7283....0F} contain approximately two million sources, while more than 65\% of them are quasar candidates. In version 7.2 \citep[v7.2 ][]{flesch2021million}, about 0.57 million candidates were removed because they did not have any radio/X-ray/WISE associations and are unlikely to be targeted in the decade. The final version (v8) removes some sources with parallaxes or proper motions detected in $Gaia$, and retains only those candidates with pQSO \textgreater 99\%, thus dramatically reducing the number of candidates. We combine the v7.2 and v8 catalogs, removing non-quasar sources already confirmed in them (e.g., accurately identified as stars in Milliquas) to complement our input catalog. This ensures that our input catalog contains the most recent quasars, while retaining as many reliable candidates as possible.

The combined sample contains many duplicate sources. To address this, we cross-matched the supplemental sample with $Gaia$ DR3 using a $3^{\prime\prime}$ radius and removed the duplicates by the unique \texttt{source\_id} of $Gaia$. Additionally, we excluded sources fainter than 20.5 mag due to the difficulty of resolving them in high-resolution spectroscopic observations. Our final input catalog comprises 2,336,841 sources, of which 408,594 are spectroscopic quasars and 1,928,247 are quasar candidates.

\begin{table*}
\caption{The number of quasar (candidates) in different versions of the Milliquas catalog.}
\label{milliquas_n}      
 \centering
\begin{tabular}{c c c c c}
\hline\hline
 & V5.2 & V6.3 & V7.2 & V8 \\    
\hline                        
Type-I QSOs/AGN & 607208 & 623004& 829666&907144 \\      
Quasar candidates & 1350176&1322599&703348&66026 \\
Type-II, Bl Lac and other objects & 41080&41197&40810&48630 \\
Total& 1998464&1986800&1573824&1021800 \\
\hline
\end{tabular}
\end{table*}

\subsection{The GMP quasar catalog-I}

The GMP quasar catalog-I \citep{mannucciprep} comprises a sample of spectroscopic multiple quasars selected using the GMP technique. As described in Section \ref{introduction}, the $Gaia$ satellite is revolutionizing the field with the GMP method, which identifies multiple sources with separations between $0.15^{\prime \prime}$ and $0.8^{\prime \prime}$ thanks to the presence of multiple peaks in the light profile of $Gaia$ sources. 

In the $Gaia$ catalog, some AGN have been spectroscopically confirmed and thus possess spectroscopic redshifts obtained from surveys such as SDSS, LAMOST, 2QZ \citep{croom20092df}, and the catalog by \citet{2020MNRAS.494.3491L}, and others. However, another portion of the sample is photometrically selected and lacks known redshifts. These have been observed using the ESO Faint Object Spectrograph and Camera (v.2, EFOSC2) at the New Technology Telescope (NTT) or the DOLORES spectrograph on the Telescopio Nazionale Galileo (TNG). Results from these spectroscopic campaigns will be presented in a forthcoming publication. 

Currently, the catalog contains approximately five hundred systems with known redshifts, and they could represent physically-associated dual AGN, gravitationally-lensed systems, or AGN projected close to a foreground Galactic star.

Each source in this catalog includes at least one QSO/AGN and displays multi-component features. This catalog serves as a valuable reference, allowing us to identify similar sources from the vast number of spectroscopically unobserved objects. This significantly facilitates future follow-up observations and identifications of dual AGN and lensed systems.

\section{The selection method}
\label{s_method}
\subsection{The criteria of selection}
\label{criteria_selecting}
The GMP quasar catalog-I and Milliquas provide a large number of quasars with reliable redshifts, and based on these we can study the photometric and astrometric properties of AGN pair candidates (sources with FMP $\geq 8$), and in turn search for many similar candidates among the unidentified quasar candidates. The GMP method is excellent at searching for AGN pairs at z $\textgreater 0.5$ \citep{mannucci2022unveiling, mannucci2023gmp}, which are the objects we are most interested in. At the same time, the FMP parameter is only reliable for G $\textless 20.5$ mag, therefore sources with redshifts less than 0.5 or fainter than 20.5 mag have been removed from all samples below.

\begin{figure}
  \resizebox{\hsize}{!}{\includegraphics{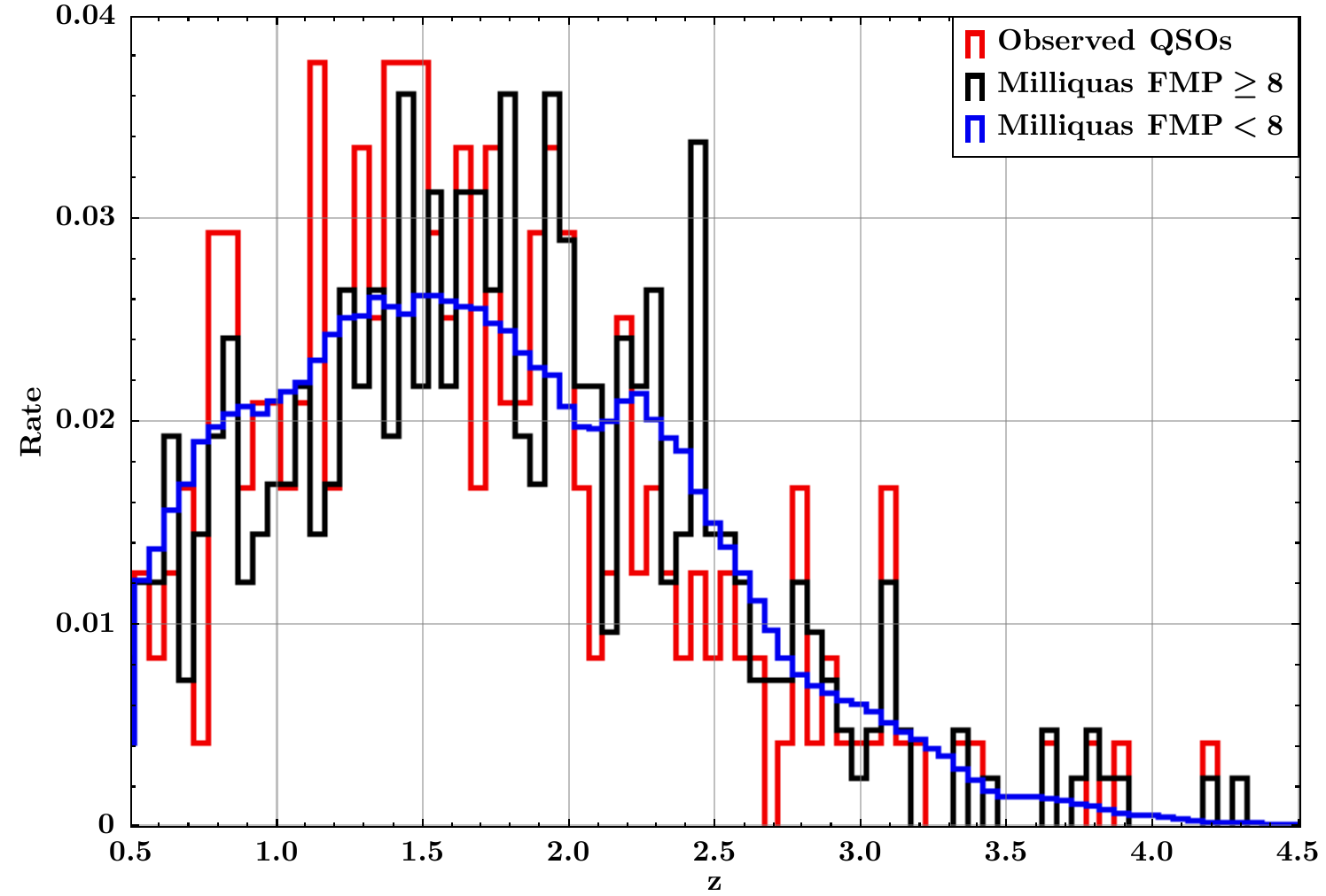}} \\
  \resizebox{\hsize}{!}{\includegraphics{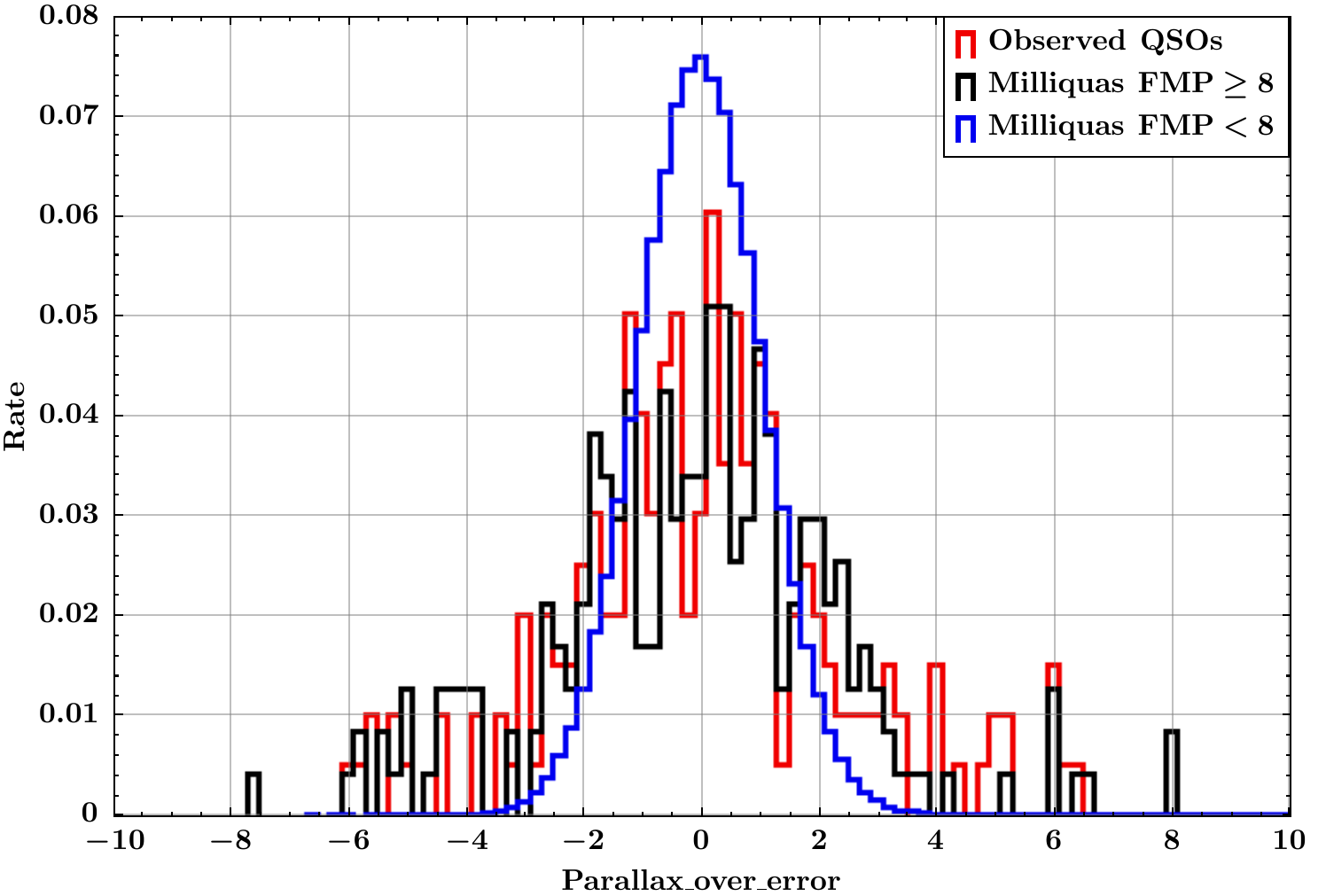}} \\
  \resizebox{\hsize}{!}{\includegraphics{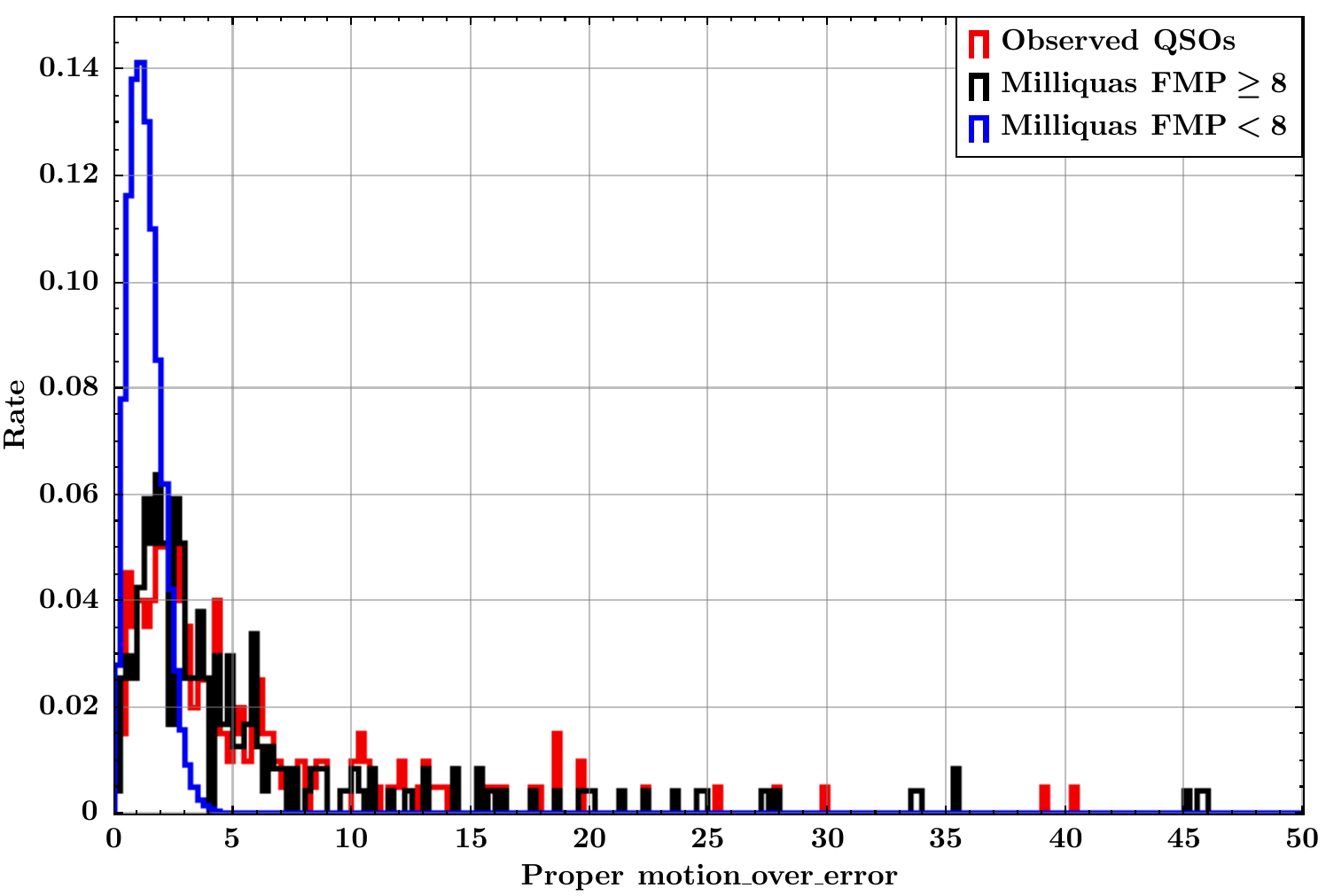}}
  \caption{Histograms of the characteristics for Observed QSOs and Milliquas. $Top$: the spectral redshift. $Middle$: the parallax\_over\_error, which is obtained by dividing parallax by parallax\_error in $Gaia$. $Bottom$: the proper motion\_over\_error, similar to the parallax\_over\_error.}
  \label{astrometry}
\end{figure}

\begin{figure}
  \resizebox{\hsize}{!}{\includegraphics{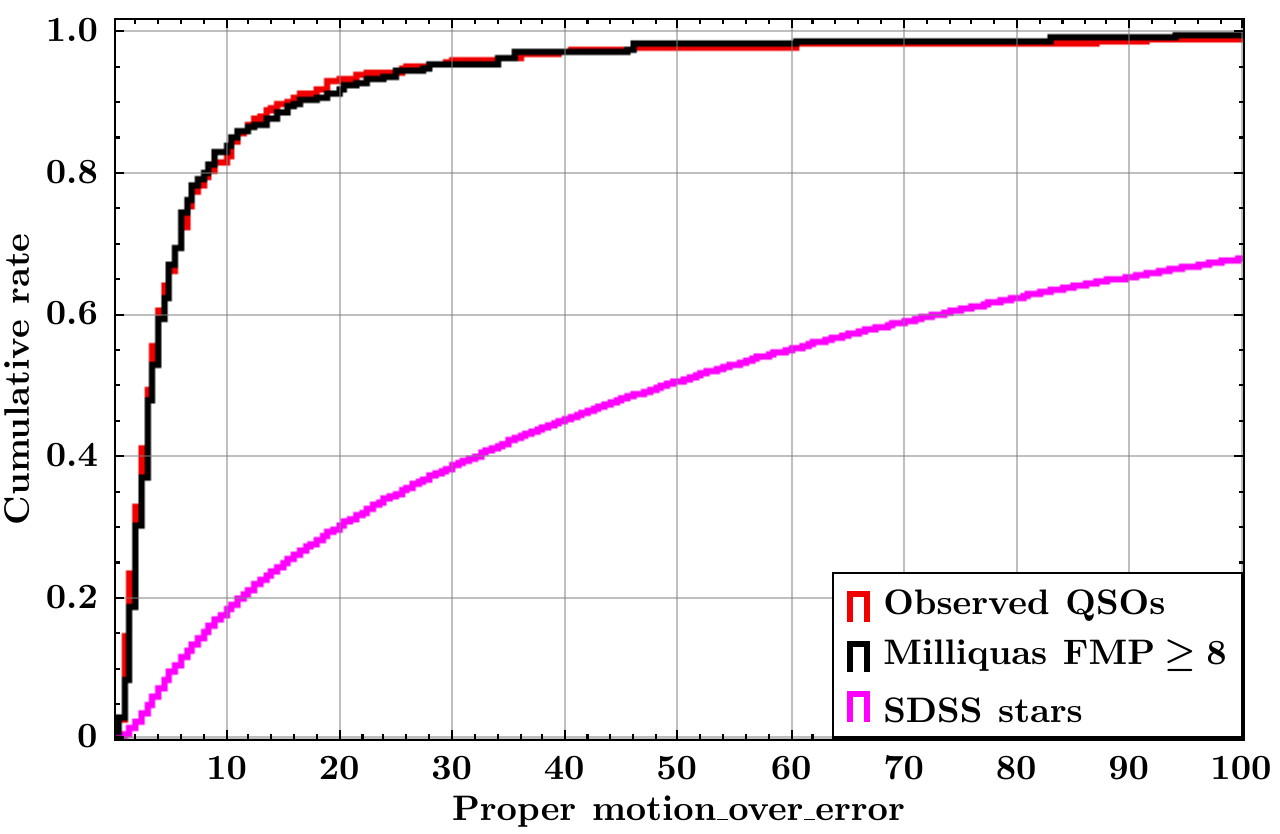}}
  \caption{Cumulative rate of normalized proper motion for different samples.}
  \label{cumu}
\end{figure}

We divided quasars (spectroscopically identified) from Milliquas into two samples based on the FMP value (cut = 8) of the sources. We then compare the redshift distributions and astrometric properties of Milliquas and Observed QSOs (the GMP quasar catalog-I, we have made follow-up observations), see Fig. \ref{astrometry}. The number of samples, designated Observed QSOs and Milliquas FMP$\geq 8$, respectively, 482 and 415. A total of 194 common sources are present in both samples. In order to accurately reflect the overall characteristics of each sample, no additional operations were performed on the common sources. After removing sources with redshifts lower than 0.5, the three samples exhibit similar redshift distributions. However, compared to the Milliquas FMP $\textless8$ sample, the other two samples show a greater dispersion in both normalized parallax and proper motion. To select pure quasars, many catalogs have been filtered using either strict parallax or proper motion criteria \citep[e.g.][]{2018A&A...618A.144G, klioner2022gaia, storey2023quaia, fu2023catnorth}. This is valid for most quasar/AGNs, but if the goal is to select a sample of AGN pair candidates with GMP $\geq8$, the previous criterion may mistakenly remove a large percentage of valuable sources. The best condition of parallax obviously is $\lvert$ Parallax\_over\_error$\rvert$ $\leq6$ in Fig. \ref{astrometry}.
To determine the optimal proper motion condition, we randomly select 300,000 reliable SDSS stars, and the cumulative normalised proper motion ratios for the different samples were plotted as in Fig. \ref{cumu}. It was found that setting Proper motion\_over\_error < 10 was able to introduce only 17\% of the stars while retaining 83\% of the GMP AGN candidates. When the normalised proper motion greater than 10, the cumulative GMP AGN candidates increases at a slow rate, while the number of stars increases rapidly, resulting in a significant increase in stellar contamination. Additionally, we investigated 15 AGNs with Proper motion\_over\_error > 30, and found that the primary reason for the observed excess proper motion is the presence of neighbor bright stars and the extened structures of sources. For this project, we chose $\lvert$ Proper motion\_over\_error$\rvert$ $\leq10$ as proper motion filter\footnote{In this paper, Parallax\_over\_error is calculated by the formula $(parallax+0.017)/parallax\_error$, where the global parallax zero point is $-0.017$ mas from \citet{brown2021gaia}. The Proper motion\_over\_error is generated using the query \texttt{$sqrt(( power(pmra/pmra\_error,2)+power(pmdec/pmdec\_error,2)-2*pmra\_pmdec\_corr*pmra/pmra\_error*pmdec/pmdec\_error)/(1-power(pmra\_pmdec\_corr,2)))$}.\label{normalize}}.

\begin{figure}
  \resizebox{\hsize}{!}{\includegraphics{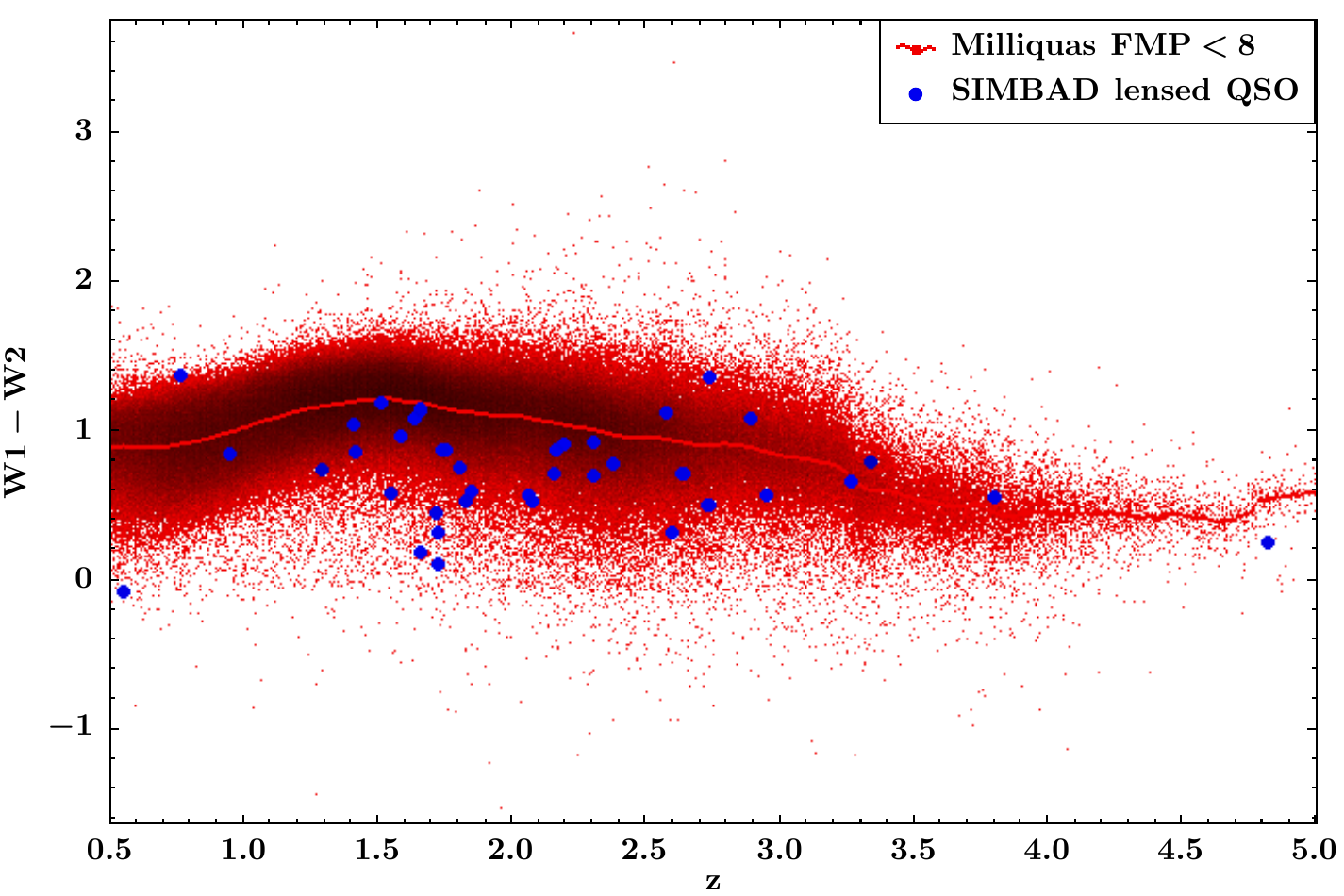}}
  \caption{Comparison of mid-infrared colors of normal AGN (FMP $\textless 8$) with SIMBAD lensed QSO at different redshifts. The red line indicates the mean IR colour of the normal AGN in each bin.}
  \label{lensed_qso}
\end{figure}

\begin{figure*}
  \resizebox{\hsize}{!}{\includegraphics{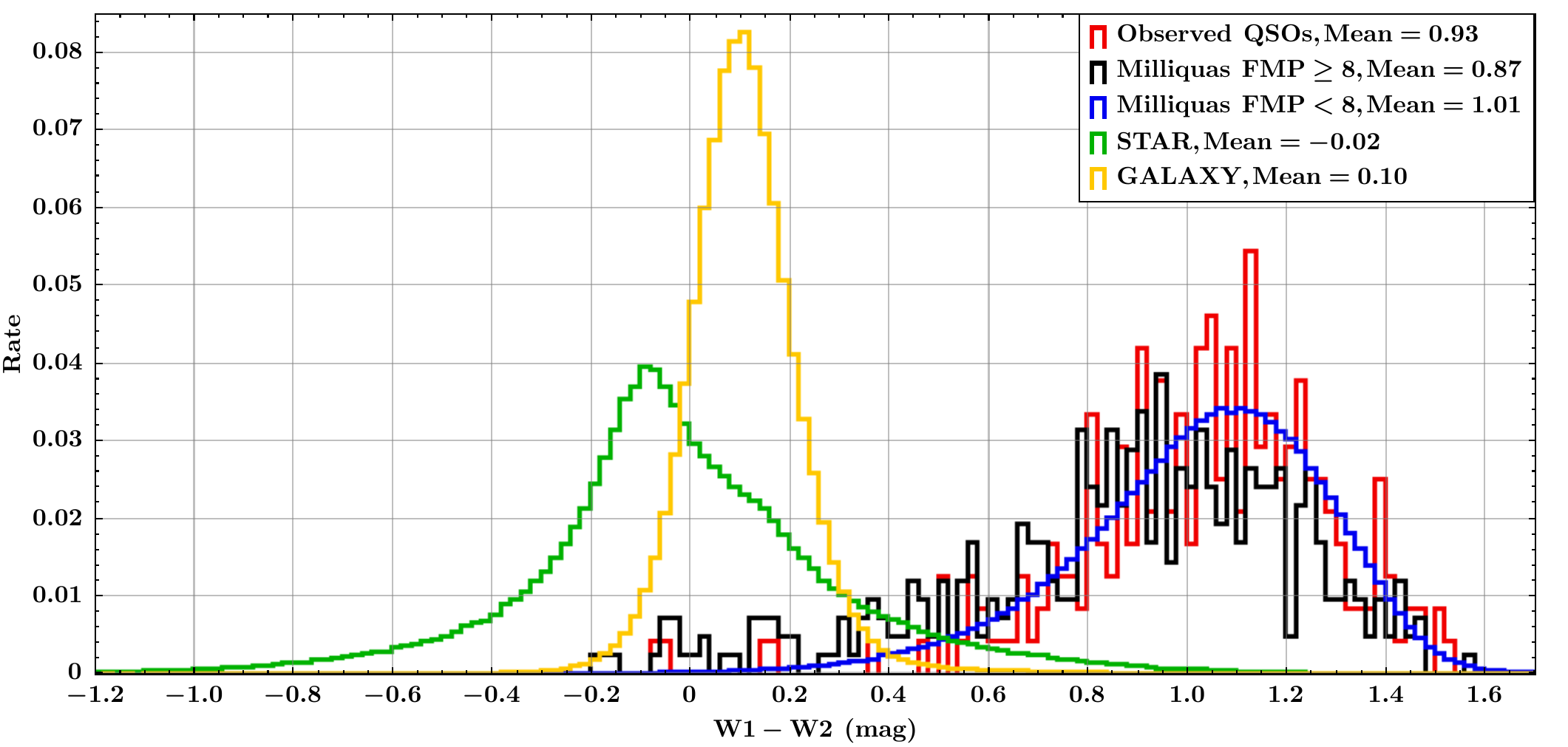}}
  \caption{Distribution of mid-infrared colors for different samples.}
  \label{wisecolor}
\end{figure*}

In general, AGN will be redder than stars and galaxies in the mid-infrared band as its accretion disk heats the surrounding dust. Therefore, the color in the mid-infrared space becomes an important criterion to effectively distinguish AGN from other sources \citep[e.g.][]{2012ApJ...753...30S, 2012MNRAS.426.3271M, 2015ApJS..221...12S, 2019MNRAS.489.4741S, storey2023quaia}. We have noted that Type II dual AGN would be bluer at the mid-infrared band \citep{2021AJ....162..276Z}, which may be attributed to the observation of Type II AGN are more susceptible to significant obscuration or contamination from host galaxies \citep{mateos2013uncovering}. And for the lensed QSO, we collect the lensed QSO with accurate redshift and WISE colors from SIMBAD database \citep{2000AAS..143....9W}. We then compared these sources with normal AGN (FMP $\textless 8$) in Milliquas, as in Fig. \ref{lensed_qso}. Most of the lensed QSO are below the average IR color of normal AGN, suggesting that the lensed QSO would be bluer. The reason could be contamination from the foreground lensing galaxies or the limited resolution of WISE observations. In summary, it is anticipated that AGN pairs will exhibit bluer infrared colors than normal AGN. To further probe the mid-infrared color distribution of GMP AGN pair candidates, we randomly selected 300,000 stars and 300,000 galaxies from the clean spectroscopy of SDSS DR18 \citep{almeida2023eighteenth} and matched all the samples with CatWISE within a radius of $3^{\prime \prime}$. Fig. \ref{wisecolor} shows the distribution of the mid-infrared color for five samples. The W1-W2 colors criterion still effectively distinguish AGNs from stars and galaxies. However, the Observed QSOs and Milliquas FMP $\geq8$ samples have lower W1-W2 values compared to the Milliquas FMP $\textless8$ sample. Therefore, if we use an extreme criterion (e.g. W1-W2 $\textgreater$ 0.8 mag), we will obtain a pure sample while losing more than 30\% of the AGN pair candidates. As a result, in order to preserve as many AGN pair candidates as possible without significantly increase the contamination, we choose W1-W2 $\geq$ 0.4 mag as the mid-infrared color criterion, similar to what was already done for the QUAIA catalog \citep{storey2023quaia}. 

\begin{figure*}
\centering
  \resizebox{0.8\hsize}{!}{\includegraphics{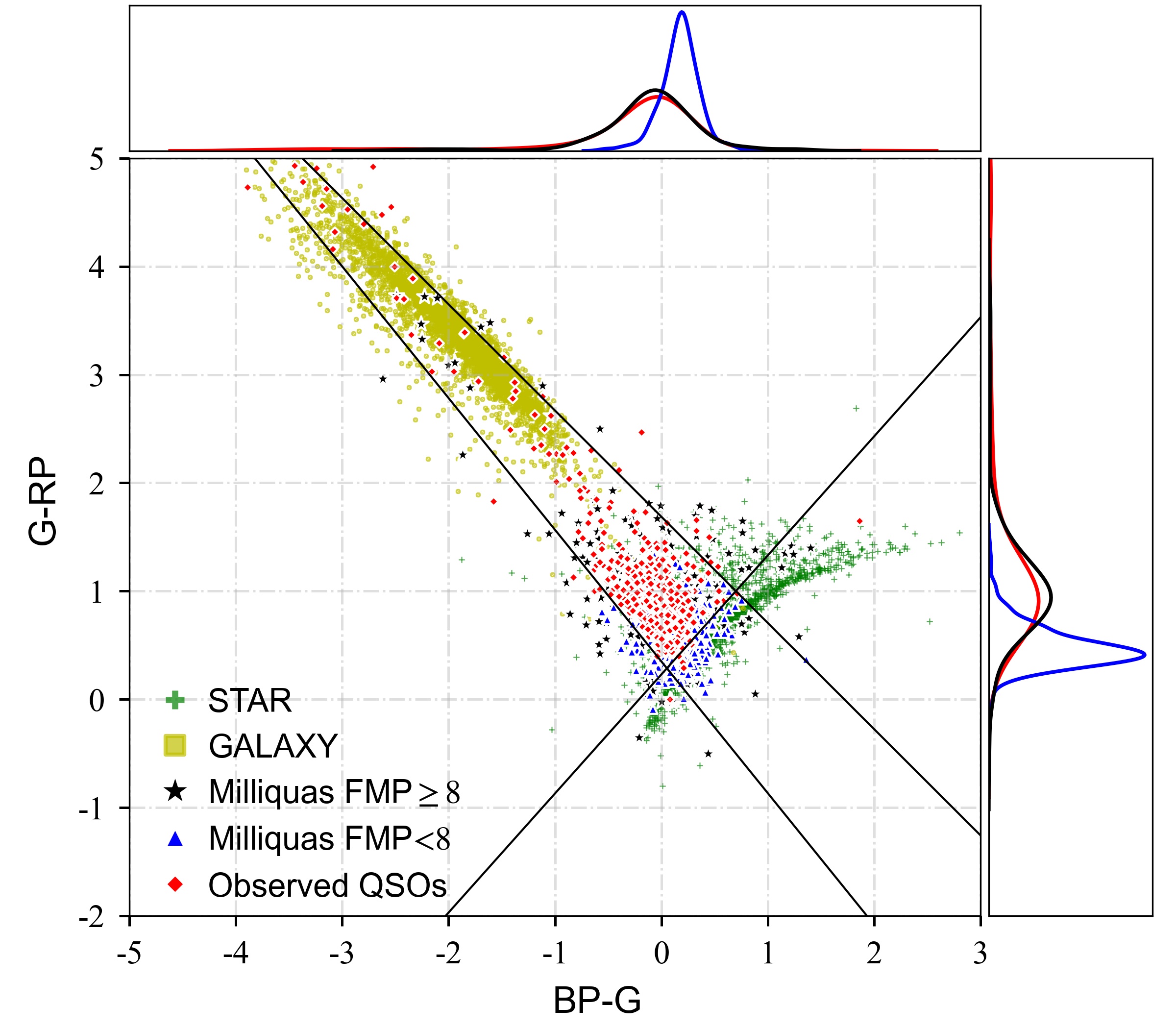}}
  \caption{Color–color diagrams of $Gaia$ colors for different samples. The distributions at the edges are plotted only for Observed QSOs and Milliquas. The black lines present the cut we used.}
  \label{gaiacolor}
\end{figure*}

Besides astrometric data, $Gaia$ DR3 also offers the BP and RP colors covering 400-500 nm, and 600-750 nm respectively. Although the gap between these two bands is not great, the difference of these two colors has been proven to be a valid criterion for selecting AGN \citep[e.g.][]{bailer2019quasar,storey2023quaia}. Fig. \ref{gaiacolor} shows the color–color diagrams of $Gaia$ BP-G and G-RP. The stars and galaxies are the same as the samples in Fig. \ref{wisecolor}. We find that both Observed QSOs and Milliquas GMP $\geq8$ have different color distributions from the Milliquas GMP $\textless 8$ sample, and some AGN pair candidates even fall into the region of galaxies. This result essentially corroborates the hypothesis that GMP AGN pair candidates are not as red as typical single AGNs, as previously suggested based on the WISE colors. We adopt the following criteria:
   \begin{equation}
   \begin{split}
      &G - RP - 1.10(BP - G) \textgreater 0.23; \\
      &G - RP + 1.22(BP - G) \textgreater 0.35; \\
      &G - RP + 0.98(BP - G) \textless 1.69. 
   \end{split}
   \end{equation}

As illustrated in Fig. \ref{gaiacolor}, the majority of AGN pair candidates have been retained, while a significant proportion of the stars, some single quasars and galaxies have been removed. These criteria would make our sample have rare stellar contamination, but would introduce a large number of galaxies. Fortunately, as shown in Fig. \ref{wisecolor}, the contamination of galaxies will be almost eliminated when W1-W2 $\geq$ 0.4 mag,  which is the mid-infrared color criterion we used.

\subsection{Selection Process}
\label{s_process}

\begin{figure}
  \resizebox{\hsize}{!}{\includegraphics{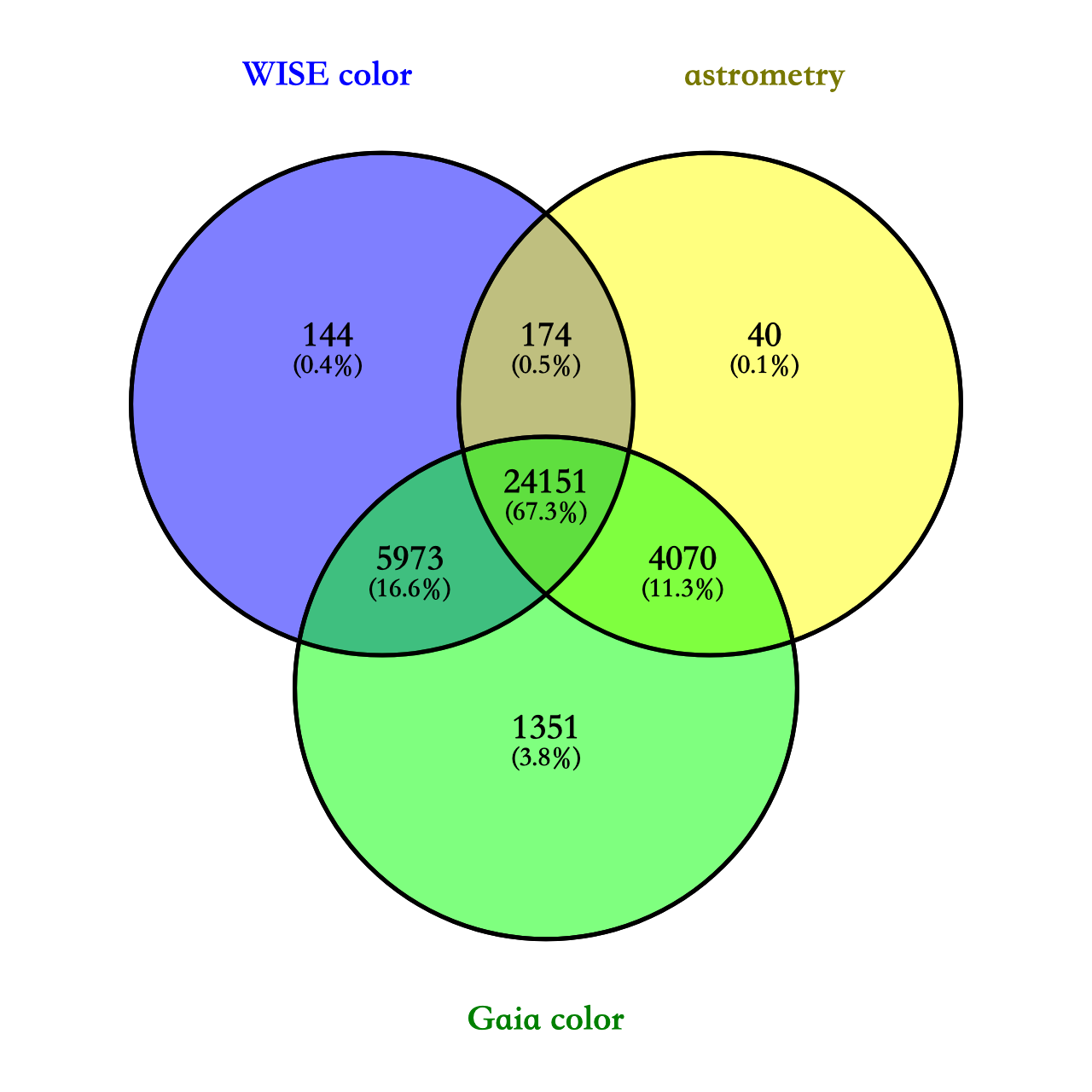}} \\
  \caption{Venn diagram of parameters available for 35,939 candidates.}
  \label{venn_parameter}
\end{figure}

\begin{figure*}
  \resizebox{\hsize}{!}{\includegraphics{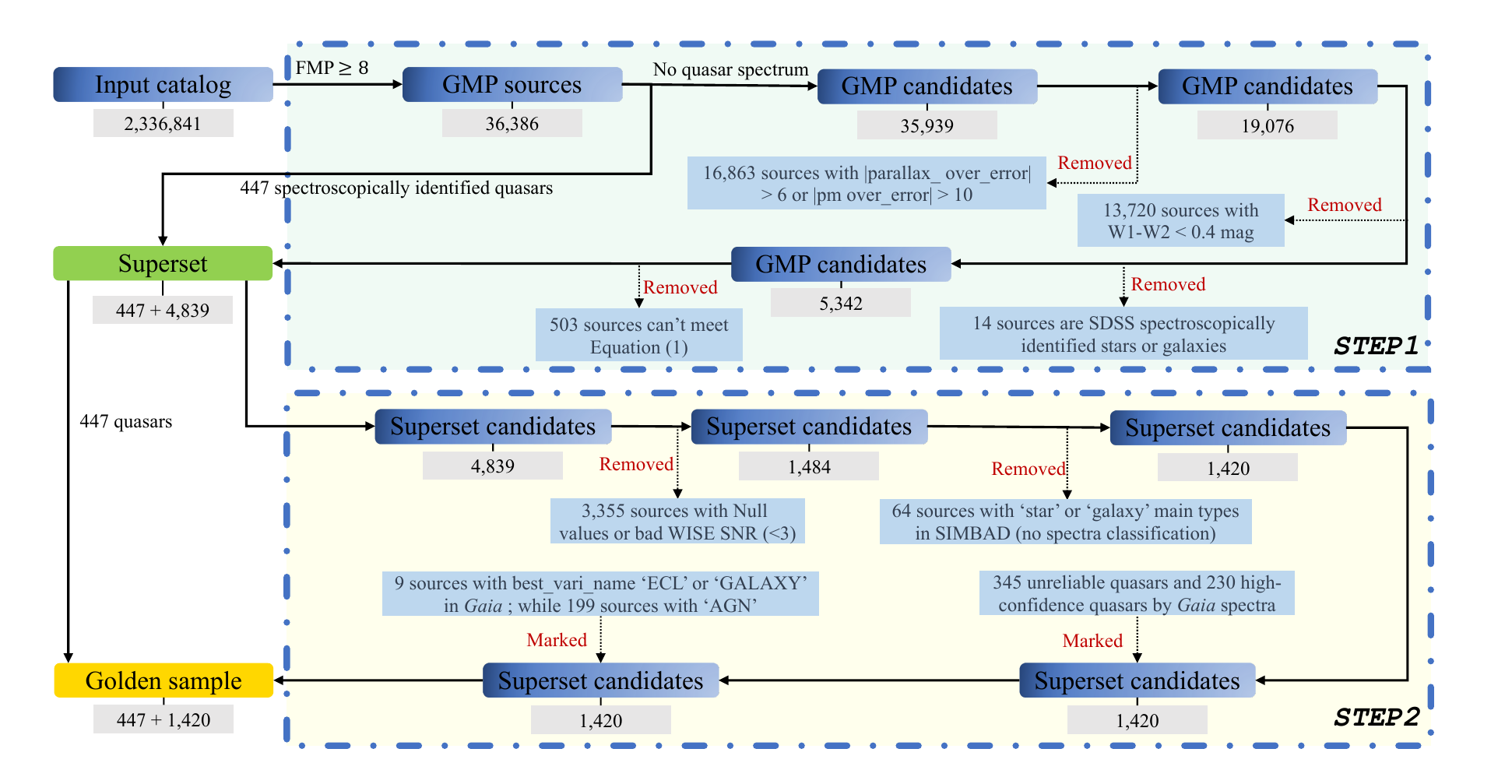}} \\
  \caption{Flowchart of the catalog selection process.}
  \label{flowchart}
\end{figure*}

In the selecting process, we first apply the GMP technique by selecting sources with FMP $\geq8$. The test on stellar pairs demonstrated that for pairs with G $\textless$ 20.5 and $\delta \textgreater 0.15^{\prime \prime}$ (the projected separation), almost all of them could be detected using the FMP $\geq8$ criterion with a very low level of contamination (below $2 \times 10^{-3}$) \citep{mannucci2023gmp}. After this step, we obtained 36,386 sources, consisting of 447 spectroscopic quasars and 35,939 candidates. In order to further purify these candidates, it is necessary to apply the criteria set out in Section \ref{criteria_selecting}. However, it has been determined that not all candidates have comprehensive color and astrometric parameters. Fig. \ref{venn_parameter} shows the availability of parameters for these sources, with only 24,151 (67.3\%) candidates having complete parameters. 

To minimise misjudgements and retain as many candidates as possible, while still providing a reliable valuable catalog to observe, we divide the selecting process into two main steps (\textit{STEP1} and \textit{STEP2}), see Fig. \ref{flowchart}. First, we sequentially removed the sources that did not meet the criteria set out in Section \ref{criteria_selecting}. During this process, we also removed 14 galaxies and stars that had been spectroscopically identified in SDSS DR16 \citep{2020ApJS..249....3A}. The characteristics of these removed sources are significantly different from those of the quasar sample, which has resulted in the creation of a relatively comprehensive superset comprising 5,286 sources. Due to many sources with missing color or astrometric parameters that cannot be removed by the method in \textit{STEP1}, some contamination may occur in the superset. In order to minimize telescope time spent on quasar-star associations, it is essential to select a sample that is as pure as possible. 

In \textit{STEP2}, the sources that lacked color or astrometric parameters were firstly removed, as well as those with low signal-to-noise ratios in the infrared band. The process removed 3,355 sources, of which 1,493, 3,016, and 82 were found to be lacking in astrometry, WISE colors, and $Gaia$ colors, respectively, and 92 exhibited infrared magnitude SNR below than 3. Subsequently, the remaining sources were cross-matched with the SIMBAD database \citep{2000AAS..143....9W}, resulting in the identification of 64 sources with a MAIN TYPE of star or galaxy. The MAIN TYPE were identified by a variety of methods in the previous literatures, but no valid spectroscopic data are available. We consider these sources to be low-confidence quasars or quasar-star pairs, which are of no interest to us and therefore were removed. 

\subsection{$Gaia$ spectrum and Variability}
The above selection concerns color and astrometry, and does not encompass detailed spectral and variability data. This is due to the dearth of useful spectral and optical variation information for these sources, rendering identification impossible. While $Gaia$'s low-resolution spectra offer some assistance, they cannot be considered a valid identification tool. \citet{delchambre2023gaia} provides classification probabilities (Specmod) for quasars obtained from $Gaia$ spectral information by machine learning. However, the \href{https://gea.esac.esa.int/archive/documentation/GDR3/Data_analysis/chap_cu8par/sec_cu8par_apsis/ssec_cu8par_apsis_dsc.html}{online documentation} indicates that even when a threshold of QSO probability greater than 0.9 is applied, the resulting purity is only about 60\%, while 100\% completeness is achieved only when the probability is very close to zero. Verification of the Milliquas revealed that approximately 12\% of the spectroscopically identified quasars exhibited QSO probability values below 0.001 in $Gaia$. Consequently, we didn't removed any sources in this step; rather, 345 unreliable quasars with probabilities below 0.001 and 230 high-confidence quasars with probabilities above 0.9 were flagged. With regard to the classification of quasars using $Gaia$ light-variation information, \citet{carnerero2023gaia} and \citet{rimoldini2023gaia} demonstrated 95\% purity in their respective classifications. However, only 208 sources in our sample have light-variation information, and 199 of them are identified as AGN in \citet{rimoldini2023gaia}. The remaining 9 sources are more likely to be galaxies or eclipsing binaries. Therefore, we also annotate these high-confidence or low-confidence sources in our catalogs. Finally, we were able to obtain a sample of 1,867 sources, of which 447 were spectroscopically identified quasars and 1,420 were highly reliable GMP AGN pair candidates.

\section{Properties of the catalog}
\label{s_character}
We provide two catalogs, one is a superset with high completeness but low purity, with 5,286 sources. The other is the Golden sample, a catalog of GMP AGN pair candidates with high purity, containing 1,867 sources. In order to select sources with different priorities for follow-up observations, we perform a more detailed feature exploration of the sources in the Golden sample. 

\subsection{The neighbors and identified pairs}
\begin{figure}
  \resizebox{\hsize}{!}{\includegraphics{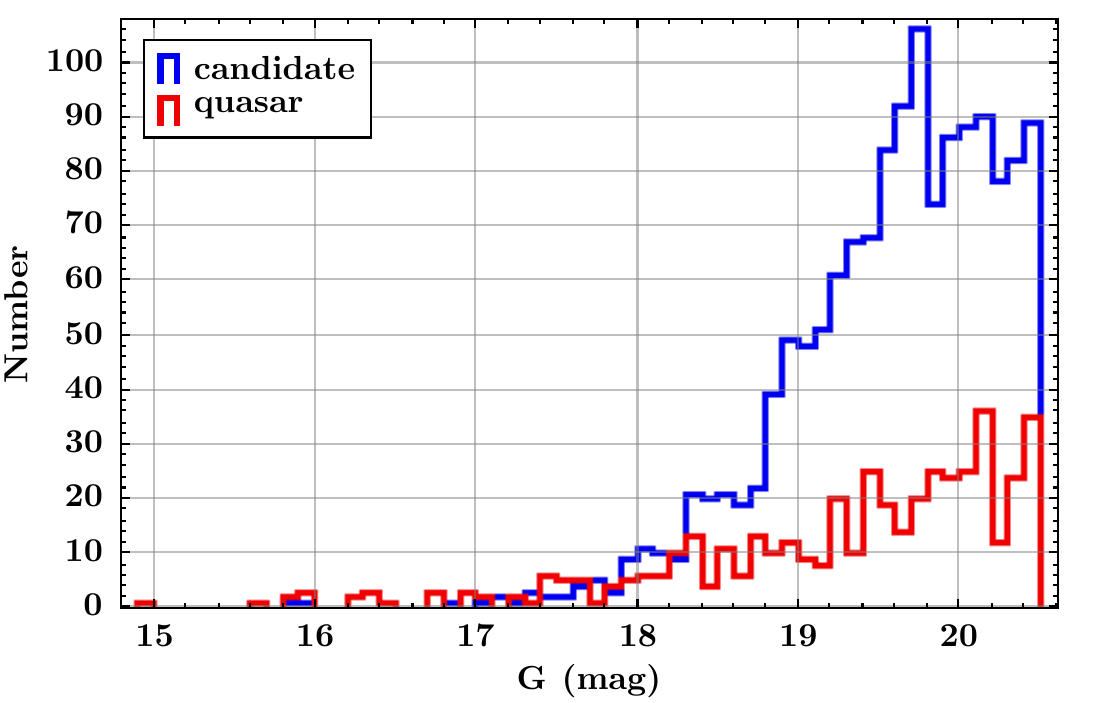}} \\
  \caption{The $Gaia$ G mag distribution of Golden sample sources.}
  \label{mag}
\end{figure}

Fig. \ref{mag} shows the magnitude distribution of the quasars and candidates in the Golden sample, with only one quasar brighter than 15.5 mag, IGR J18249-3243, a Seyfert 1 AGN \citep{masetti2009unveiling}. About 90\% of the sources are in the magnitude range of 18.5-20.5 mag, and these faint sources may have a high value of FMP due to nearby random stars, so we examined the number of nearby stars within $5^{\prime \prime}$ of the sources in the sample. The results indicate that 984 sources did not detect any other $Gaia$ sources within $5^{\prime \prime}$, 704 sources detected one neighbors, 107 sources detected two neighbors, and 72 sources detected more than two neighbors. Subsequently, an in-depth analysis was conducted to identify the underlying causes of these multiple neighbouring sources, as shown in Table \ref{multip_neighbor}. For sources with three neighbors, five of them have been identified as lensed systems, three are lensed QSO candidates, which further demonstrates the high efficiency of the GMP method for the selection of AGN pair candidates. For sources with more than three neighbors, the majority are located in dense star fields, including the SMC, LMC, the Galactic plane, and Seyfert galaxies. While these sources could be AGN pairs, the presence of dense star fields may hinder the identification, suggesting that these sources should be given a lower priority for follow-up observations. 

Furthermore, we have labelled the AGN pairs that have been identified in Golden sample by consulting the relevant literature \citep[e.g.][]{2014MNRAS.440..870T, 2018MNRAS.475.2086A, 2018MNRAS.479.5060L, 2021NatAs...5..569S}. The Golden sample comprises 39 sources that have been identified. In detail, 2, 7 and 26 out of 477 quasars have been identified as AGN-Star pairs, dual AGNs and lensed systems, and 4 out of 1420 quasar candidates identified as lensed systems. This is readily comprehensible, given that the majority of previous literature on identified AGN pairs has involved the selection of pair candidates among already identified quasars, with quasar candidates being rarely considered. Consequently, the discovery rate of pairs in  Golden sample quasars is considerably higher than in candidates, which is a selection effect. In addition, the relationship between these identified pairs and the neighbors of $Gaia$ sources was investigated. See Table \ref{corr}, the distinct distribution patterns between dual AGN and lensed systems indicate that researchers studying dual AGN should focus on $Gaia$ sources with neighbors $\leq 1$, while those investigating lensed systems may find it more effective to target sources with neighbors between 1 and 3.

\begin{table*}
\caption{The detailed examination of the factors that contribute to the presence of multiple nearby sources.}
\label{multip_neighbor}      
 \centering
\begin{tabular}{c c >{\centering\arraybackslash}p{10cm}}
\hline\hline
Number of neighbors & Count of Golden sample sources & Reasons \\    
\hline                        
3 & 31&13 near Galactic plane, 5 are lensed systems, 3 are lensed candidates1 near NGC 5466, 3 near NGC5272 , 5 near LMC and 1 near SMC \\      
4 & 15&11 near Galactic plane, 1 is Seyfert\_2, 2 near LMC and 1 near SMC \\
5 & 9&6 near Galactic plane and 3 near LMC \\
6& 8& 4 near Galactic plane, 2 are Seyfert\_2, 1 near LMC and 1 near SMC  \\
7 & 4& 2 near LMC, 1 near SMC and 1 near Galactic plane \\
10 & 3& 3 near LMC \\
13 & 1& 1 near NGC 1261 \\
14 & 1& 1 is NGC 986 Seyfert\_1 \\
\hline
\end{tabular}
\end{table*}

\begin{table}
\caption{Correlation of AGN pairs with $Gaia$ neighbors in a radius of $5^{\prime \prime}$. These 39 pairs presented here have been identified in the  literature and belonging to the Golden sample.}
\label{corr}      
 \centering
\begin{tabular}{c|c|c|c}
\hline\hline
Category & AGN-Star & Dual AGN & Lensed system \\    
\hline                        
single source  & 1 & 4& 5 \\
\hline 
one neighbor  & 1&3&11 \\
\hline 
two neighbors & -&-&9 \\
\hline 
three neighbors& -&-&5 \\
\hline
\end{tabular}
\end{table}

\subsection{The redshift}

Milliquas provides high-precision spectroscopic redshifts for the majority of quasars. However, there are still more than one thousand sources in the Golden sample that do not have valid spectroscopic redshifts. $Gaia$ provides redshifts derived from low-resolution BP/RP spectra, which are subject to great errors and systematic offsets \citep{delchambre2023gaia}. Recently, \citet{storey2023quaia} (quaia) and \citet{fu2023catnorth} (CatNorth) have improved the redshifts of some $Gaia$ sources with photometric data from other telescopes, thereby substantially reducing the systematic errors. Although the error is significant, photometric redshift can provide a reference redshift for a large number of sources without spectroscopic observations. Moreover, a lot of researchs have demonstrated that machine learning has a highly effective capacity for inferring photometric redshifts \citep[e.g.][]{curran2021qso, razim2021improving, hong2023photoredshift}. \citet{2022MNRAS.516.2662K} has inferred photometric redshifts of 4.8 million quasar (candidates) using the colours of Pan-STARRS \citep[PS1,][]{chambers2016pan} and WISE \citep{wright2010wide}, with mean squared error of $\sim$0.2. 

The majority of the redshift in our Golden sample is derived from the above literature, with the highest priority given to Milliquas. For sources that did not provide redshifts in Milliquas, we supplemented them from QUAIA with CatNorth. The remaining sources that still did not have redshift data were supplemented in $Gaia$ and PS1-WISE. The sources of redshifts for all sources are shown in Table \ref{z_reference}. A total of 298 sources in the Golden sample lack redshift. The majority of these sources lack spectroscopic information from $Gaia$ as well as multi-band photometric information to infer the redshift. We note that the redshifts given by references other than Milliquas are derived wholly or partially from photometric information, and are therefore less precise. In some cases, these may even have large variations from the true redshift. Nevertheless, these redshift data can serve as a valuable reference for researchers, enabling them to identify and filter sources within their desired redshift range.

\begin{figure*}
\centering
\includegraphics[width=0.3\hsize]{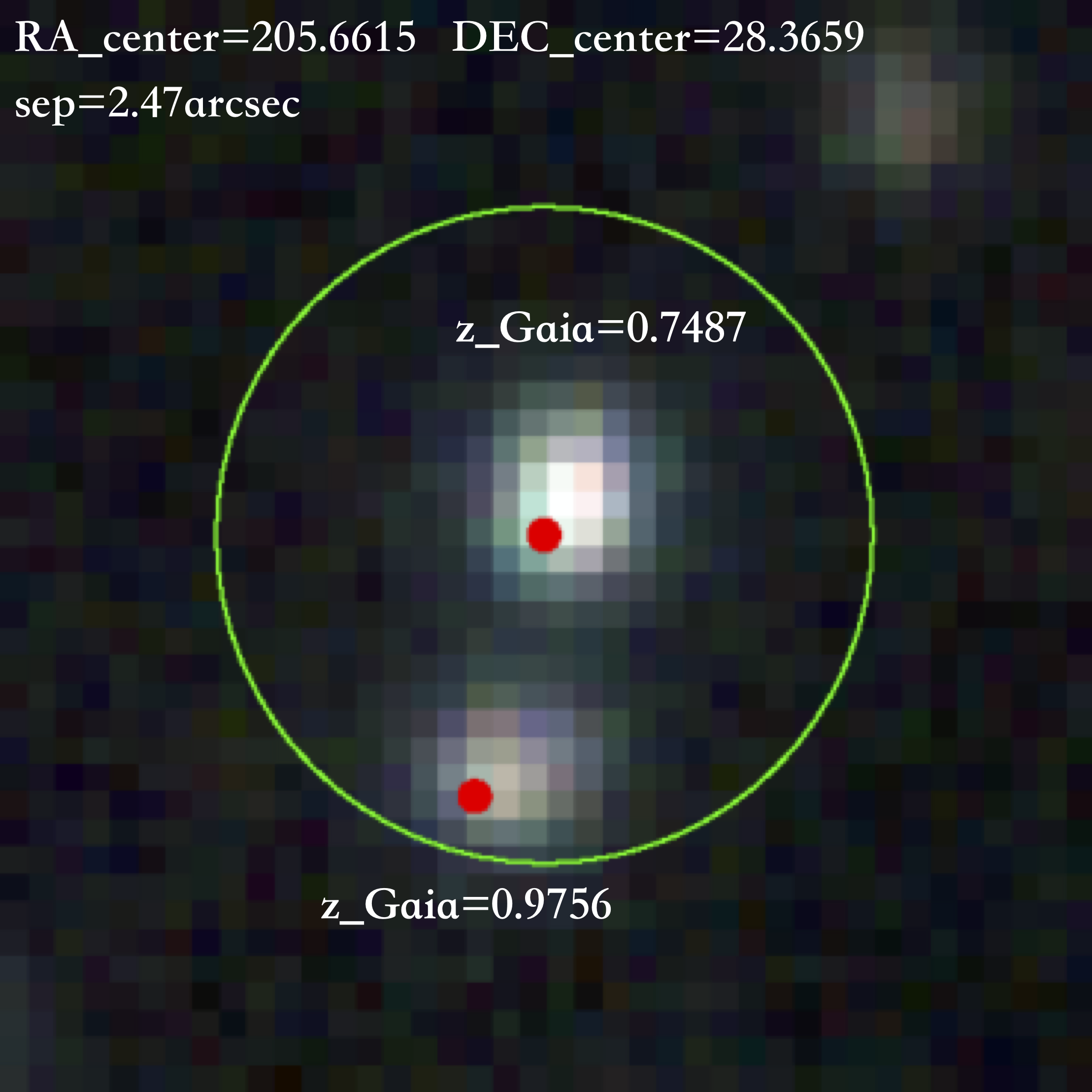}
\includegraphics[width=0.3\hsize]{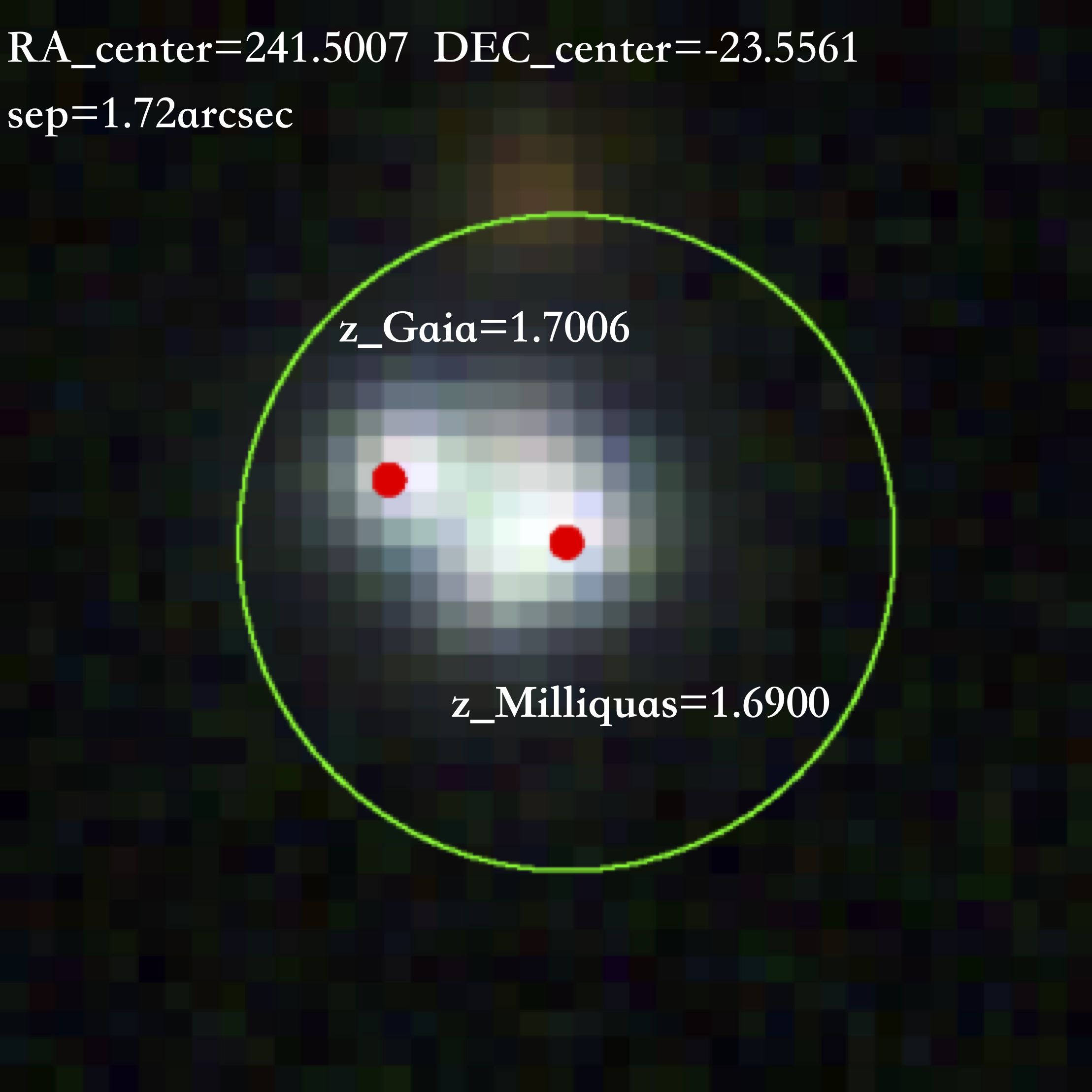}
\includegraphics[width=0.3\hsize]{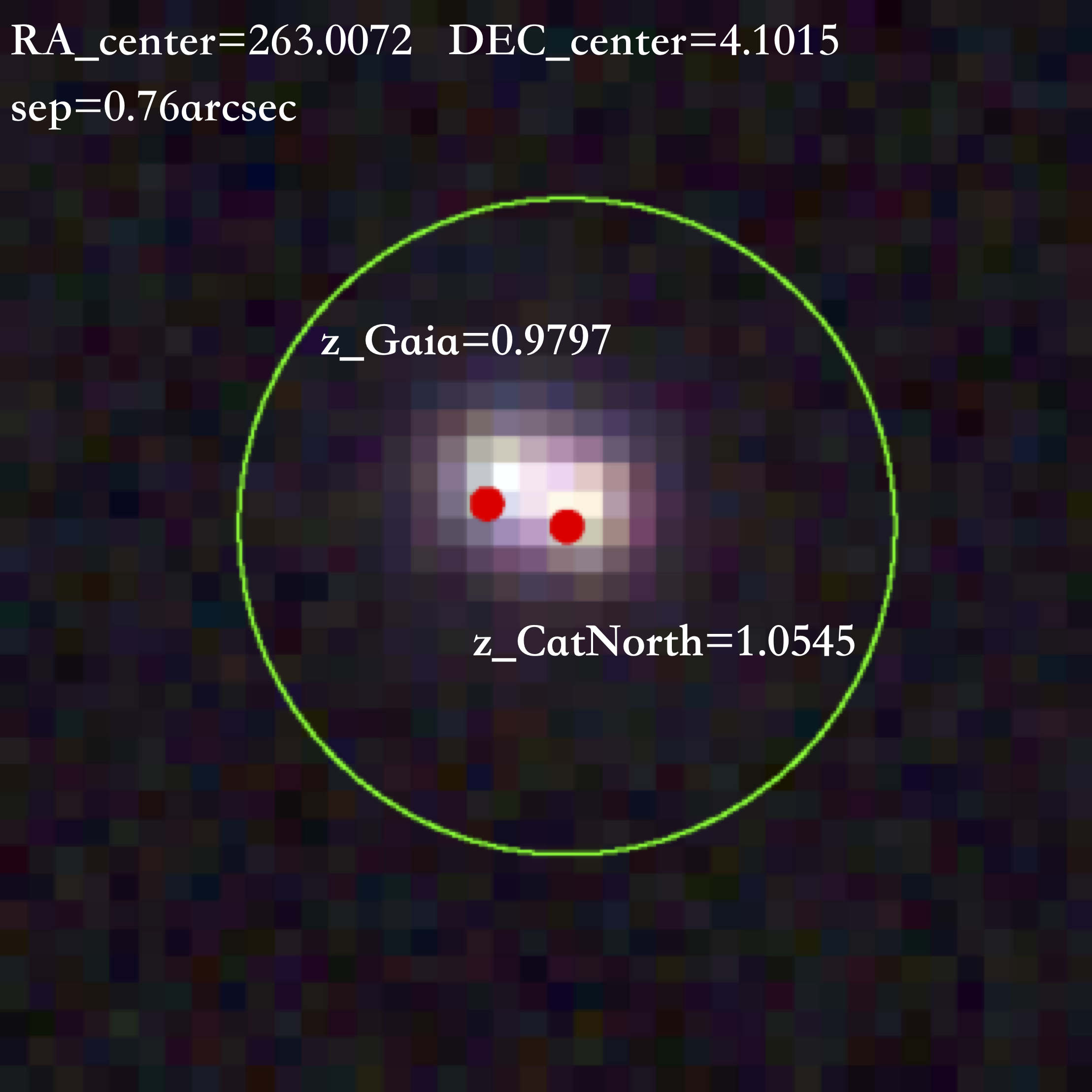}
    \caption{PS1 composite colour images of three pairs in the Golden sample. The images were obtained by overlaying the i, r, and g bands of PS1, which were set to red, green, and blue, respectively. The size of each cutout is $10^{\prime \prime} \times 10^{\prime \prime}$. The coordinates of the image centers, the projection separations of the components, and the reference redshift of each component are indicated in the images. The red dots represent the coordinates of the components in $Gaia$, and the green circles have a radius of $3^{\prime \prime}$.}
        \label{three_sources}
\end{figure*}

It is relatively straightforward to identify pairs of sources within the Golden sample. For instance, by setting a radius of $3^{\prime \prime}$ and examining the results, we find three pairs of sources that have similar redshifts. This is illustrated in Fig. \ref{three_sources}. The largest difference in redshift among these three pairs reaches 0.23, as shown in the left panel of Fig. \ref{three_sources}. However, the component with a redshift of 0.9756 is labelled as a poorer spectroscopic observation in $Gaia$, which could result in significant errors in the obtained redshift. We consider all three pairs to be high-probability dual or lensed AGNs. It is noteworthy that these three pairs exhibit disparate separations, which are 2.47, 1.72, and 0.76 arcsec. This illustrates the efficacy of the GMP method in screening pairs with varying separations. Upon cross-matching with previous literature, the second pairs (RA\_center=241.5007, DEC\_center=-23.5561) was identified as a lensed AGN by \citet{2018MNRAS.479.5060L}. Furthermore, it is important to acknowledge that the redshift from $Gaia$ BP and RP may not be entirely reliable, particularly when the two sources are close. In such cases, the other two pairs (see left and right panel of Fig. \ref{three_sources}) may necessitate more detailed observations to determine their natural characteristics.

\begin{table}
\caption{The reference catalogs of the Golden sample redshift.}
\label{z_reference}      
 \centering
\begin{tabular}{c|c}
\hline\hline
Redshift reference & Number of sources \\    
\hline                        
Milliquas& 471 \\
\hline 
quaia&326 \\
\hline 
CatNorth&153 \\
\hline 
$Gaia$&491\\
\hline 
PS1-WISE&128\\
\hline
No data&298\\
\hline
\end{tabular}
\end{table}

\section{Conclusions and the future}
\label{s_conclusion}
As a new approach to select AGN pairs at sub-arcsec separations, the GMP method employs the parameter \texttt{ipd\_frac\_multi\_peak} in $Gaia$ DR3. A series of previous studies has demonstrated the efficacy of this method in selecting pairs with separations of $0.15^{\prime \prime}-0.8^{\prime \prime}$. However, the sources utilized for selecting GMP systems are predominantly drawn from AGN that have been spectroscopically identified by large surveys (e.g. SDSS, LAMOST), with a limited number of exceptions. This study aimed to investigate the differences between single AGN and AGN pairs in astrometric, multi-band colours based on the Milliquas and GMP quasar catalog-I. By analysing these differences, we were able to develop a more accurate method for selecting reliable AGN pair candidates from over 6 million quasar candidates, significantly expanding the sample size. The main conclusions of this study are as follows:
\begin{enumerate}
    \item In cases where the redshift distributions are identical, GMP AGNs show more significant parallaxes and proper motions than non-GMP AGNs. Concurrently, the mean value of the colour in the infrared band of these AGN is also bluer than that of normal AGN. Furthermore, the optical band colour observed by $Gaia$ exhibits a markedly disparate distribution between the two types of samples.
    \item We conducted a study to identify the best criteria for selecting AGN pair candidates. The majority of sources in our input catalog lacked spectroscopic information, so we employed astrometric and colour information screening to obtain a complete superset containing 5,286 sources. We then proceeded to obtain a pure Golden sample of 1,867 sources through further rigorous selection and a match of the previous literature.
    \item In order to prioritise candidate sources for the follow-up observations, some sources were marked as high-reliability or low-reliability in the Golden sample. This classification was based on information regarding light variations as well as low-resolution spectra obtained from $Gaia$. Some Golden sources with  excessive neighbouring sources were examined, and AGN pairs identified in the previous literature were labelled. We found 2, 7, and 30 sources identified as AGN-Star, dual AGN, and lensed AGN, respectively.
    \item The majority of sources in the Golden sample have been provided with redshift information through the integration of multiple catalogs. It was found that three pairs of sources in the Golden sample had separations of less than $3^{\prime \prime}$ and similar redshifts between components. One of the pairs has been identified as a lensed AGN, while the other two lack valid spectroscopic observations. Concurrently, the angular distances between these three pairs are 2.47, 1.72, and 0.76 arcsec, respectively. This indicates that the GMP method can be employed to select effectively not only for close pairs ($0.15^{\prime \prime}-0.8^{\prime \prime}$), but also for pairs with larger separation.
\end{enumerate}

This is the first attempt to select GMP pair candidates from over six million sources, and we ultimately identify about 1,800 high-value sources, of which over 1,500 provide reference redshifts. It is anticipated that comprehensive observations of this sample will yield a substantial number of AGN pairs for the community. In the process of selection, in order to avoid excessive contamination, we removed a number of sources located in the dense star fields, which would directly lead to noticeable deficiencies in our sample. It is also important to note that hundreds of sources in our Golden sample have been labelled as unreliable quasars. These sources may exhibit characteristics that differ from the AGN population due to poor data quality. Nevertheless, we maintain the view that a significant proportion of these sources are still real AGN pairs, which will require higher precision data to accurately identify them. 

In addition, a potential application could be in the area of celestial reference frames, where it was found that approximately eight hundred sources in the Golden sample were used to construct the $Gaia$-CRF3 \citep{klioner2022gaia}. Given the potential multi-peak structure of these sources, it is recommended that they be used with caution when constructing reference frames.

\begin{acknowledgements}
This work has made use of data from the European Space Agency (ESA) mission \href{https://www. cosmos.esa.int/gaia}{$Gaia$}, Wide-field Infrared Survey Explorer (WISE) and \href{www.sdss.org}{Sloan Digital Sky Survey (SDSS)}. We are also very grateful to the developers of the TOPCAT \citep{taylor2005topcat} software, the \href{https://simbad.cds.unistra.fr/simbad/}{SIMBAD astronomical database} and \href{https://healpix.sourceforge.io/}{HEALPix}. This work has been supported by the Strategic Priority Research Program of the Chinese Academy of Sciences, Grant No.XDA0350205, the National Natural Science Foundation of China (NSFC) through grants 12173069, the Youth Innovation Promotion Association CAS with Certificate Number 2022259, the Talent Plan of Shanghai Branch, Chinese Academy of Sciences with No.CASSHB-QNPD-2023-016. Part of this work was supported by the German \emph{Deut\-sche For\-schungs\-ge\-mein\-schaft, DFG\/} project number Ts~17/2--1. We acknowledge the science research grants from the China Manned Space Project with NO. CMS-CSST-2021-A12 and NO.CMS-CSST-2021-B10.\\ This publication  was produced as part of the PhD program in Space Science and Technology at the University of Trento, Cycle XXXIX, with the support of a scholarship financed by the Ministerial Decree no. 118 of 2nd march 2023, based on the NRRP - funded by the European Union - NextGenerationEU - Mission 4 "Education and Research", Component 1 "Enhancement of the offer of educational services: from nurseries to universities” - Investment 4.1 “Extension of the number of research doctorates and innovative doctorates for public administration and cultural heritage”.\\
      
\end{acknowledgements}

%
\bibliographystyle{aa} 
\bibliography{reference} 
%
\begin{appendix} 
\section{The details of our catalogs}
In Section \ref{s_method}, we obtained two catalogs. For the superset, we provide the right ascension, declination, other astrometric parameters and colour parameters (if available). For the Golden sample, we provide additional columns, including reliability flags as mentioned in Section \ref{s_process}, and neighbouring sources and redshift information as mentioned in Section \ref{s_character}. The details of each column in the Golden sample are shown in Table \ref{details}. These two catalogs will be available to the public at the time of the publication of this paper.

\begin{figure}
  \resizebox{\hsize}{!}{\includegraphics{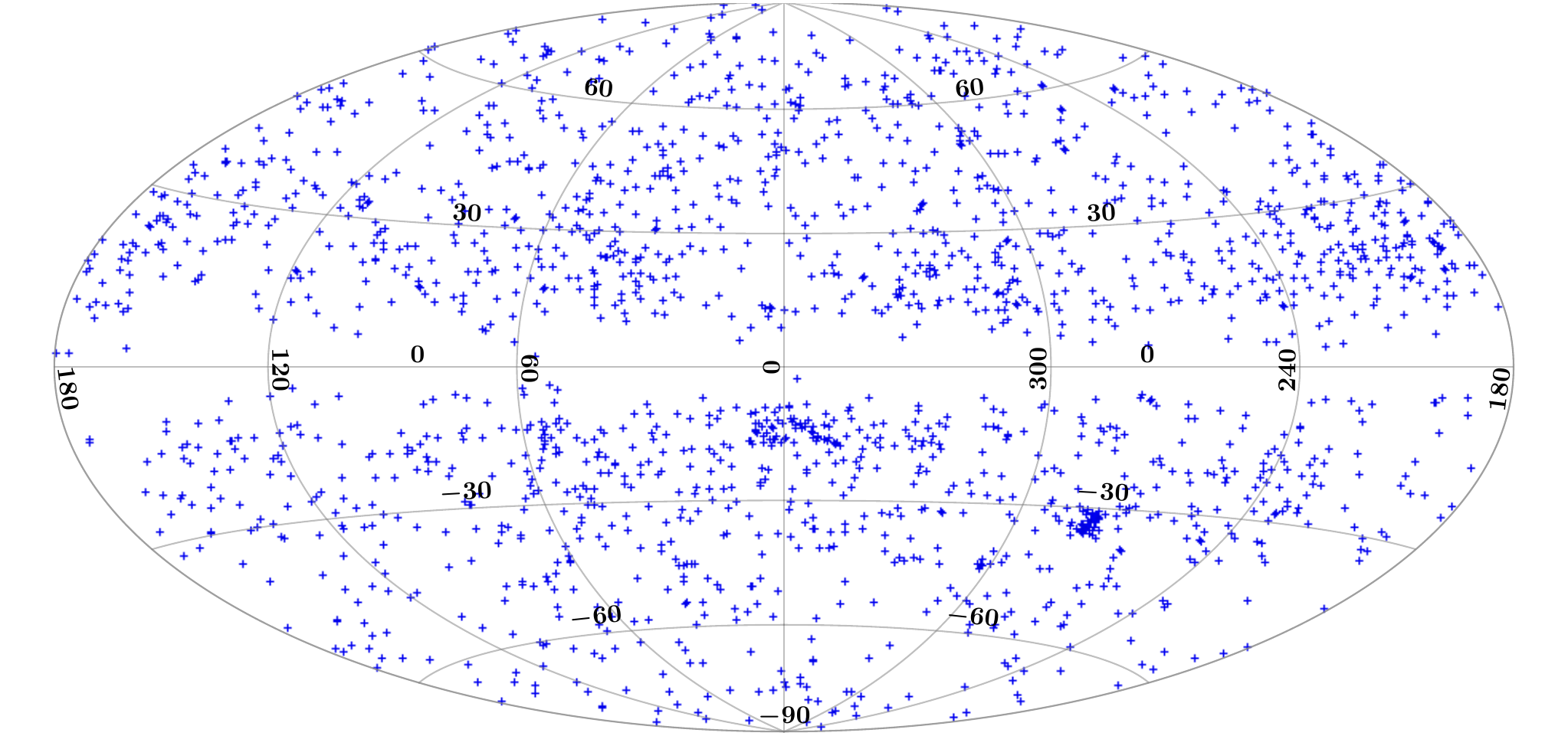}} \\
  \caption{The sky distribution of Golden sample sources, using the Hammer Aitoff projection in Galactic coordinate.}
  \label{golden_map}
\end{figure}

\begin{figure}
  \resizebox{\hsize}{!}{\includegraphics{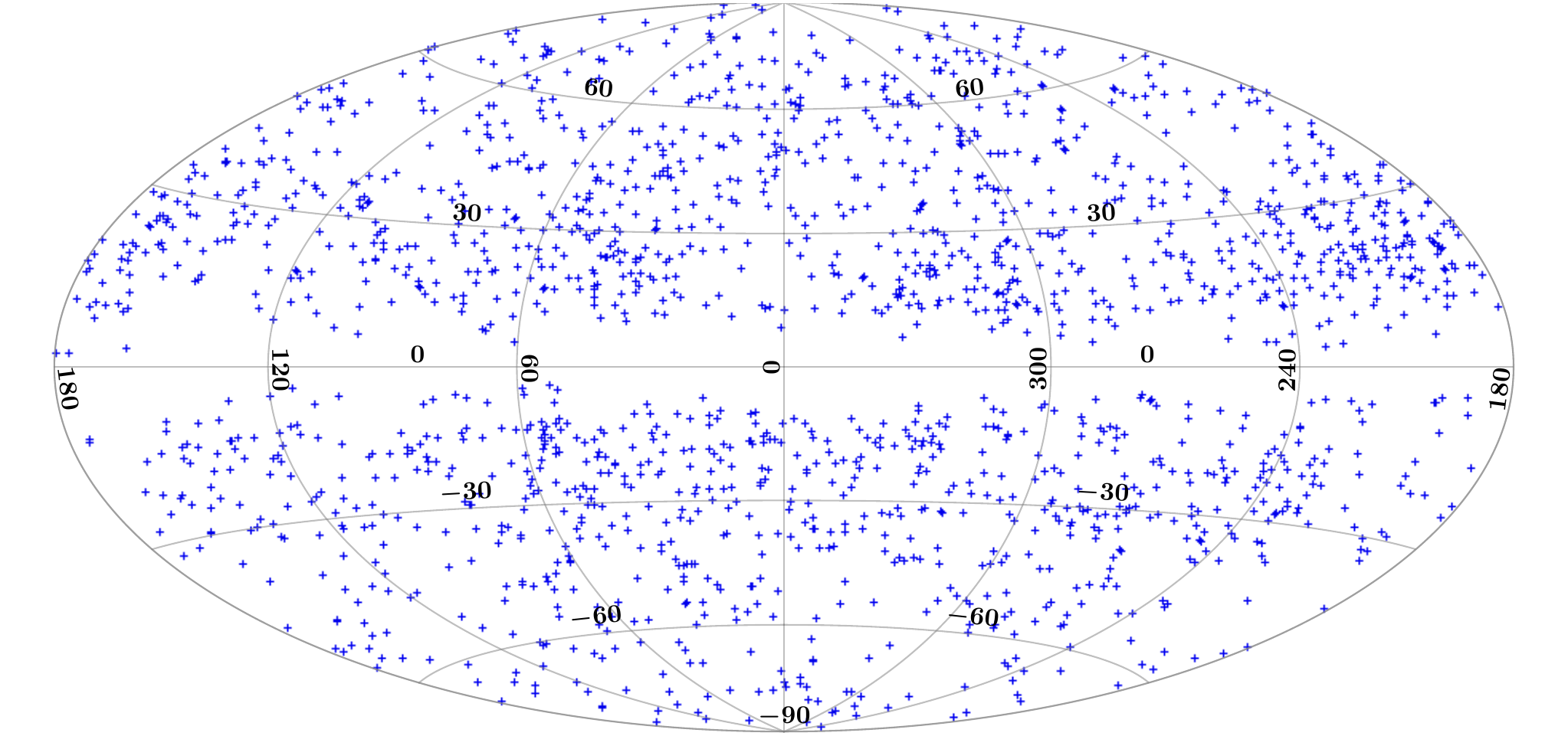}} \\
  \caption{The sky distribution of Golden sample sources without the dense regions, using the Hammer Aitoff projection in Galactic coordinate.}
  \label{golden_clear}
\end{figure}

The distribution of the 1,867 sources in the Golden sample is illustrated in Fig. \ref{golden_map}. The ideal distribution of dual AGN or lensed QSO on the celestial sphere would be uniform. However, the regions of Galactic center and the LMC in Fig. \ref{golden_map} exhibit an excess of these sources. The QSOs near the LMC are almost from a study dedicated to investigating quasars behind the Magellanic Clouds \citep{2013ApJ...775...92K}. The high stellar density of the Galactic center and LMC would suggest that these sources are more likely to have a higher FMP (see Fig. \ref{gmp8_prob}). These sources have a high probability of being AGN-Star pairs, so we have labelled them in the Golden sample. People may choose to eliminate these sources using the condition \\ \texttt{select * from Golden sample \\ WHERE neighbor\_source\_comment != `near Galactic plane' \\ \& neighbor\_source\_comment != `near LMC' \\ \& neighbor\_source\_comment != `near SMC'}. \\ This would result in a catalog of 1,769 sources distributed almost uniformly on the celestial sphere, see Fig. \ref{golden_clear}.

Furthermore, spectroscopic observations of some AGN candidates in the Golden sample have recently been conducted, with the preliminary results presented in Table \ref{identify}. On the basis of the spectral data, it was established that the 12 primary sources are AGN, while one source with a low signal-to-noise ratio (S/N) is likely to be a star. The decomposition of these sources is currently underway, and the final results will be presented in subsequent papers.

\begin{table*}
\caption{Detailed columns of the Golden sample}             
\label{details}      
\centering          
\begin{tabular}{c c c c c}     
\hline\hline       
Column & Name & Unit &Description& Example value\\
\hline                    
   1 & source\_id & - & $Gaia$ DR3 unique id & 2308488679004045056 \\  
   2 & ra & degree & right ascension in $Gaia$, epoch 2016 & 2.122490\\
   3 & dec & degree & Declination in $Gaia$, epoch 2016 & -36.923104 \\
   4 & phot\_g\_mean\_mag & mag& g-band mean magnitude in $Gaia$ &18.560379 \\
   5 & parallax\_over\_error & - & normalised parallax, see Footnote \ref{normalize}&2.064249 \\
   6 & proper motion\_over\_error & - & normalised proper mption, see Footnote \ref{normalize}&1.423420\\
   7 & bp\_g & mag & $Gaia$ BP-G color&-0.045782 \\
   8 & g\_rp & mag    & $Gaia$ G-RP color&0.922029 \\
   9 & w1\_w2 & mag    & CatWISE W1-W2 color&0.903999 \\
   10 & class & - & class of the spectroscopically identified result &q(quasar)/c(candidate)\\
   11 & ipd\_frac\_multi\_peak & - & $Gaia$ percent of source with more than one peak&12\\
   12& gaia\_spectra\_flag & - & QSO quality flag by $Gaia$ spectra&unreliable/high-confidence \\
   13& gaia\_vari\_flag & - & QSO quality flag by $Gaia$ light variation&unreliable/high-confidence \\
   14& n\_neighbor\_source & - & number of $Gaia$ neighbors with a radius of $5^{\prime \prime}$&2\\
   15& neighbor\_source\_comment & - & reasons for multiple neighbors & near LMC\\
   16& z & - & redshift& 2.860551 \\
   17& z\_reference & - & reference catalog of the redshift& quaia \\
   18& identified\_class & - & class of AGN pairs identified by literature&Dual\_AGN\\
   19& identified\_class\_bibcode &-& the literature bibcode& 2001ApJ...549L.155J \\
\hline                  
\end{tabular}
\end{table*}

\begin{table}
\caption{Classification of some primary sources by our spectroscopic observations.}
\label{identify}      
 \centering
\begin{tabular}{c|c|c}
\hline\hline
RA & DEC & Classification of the primary source\\    
\hline                        
75.5898& -37.8658 & AGN \\
\hline 
79.7329&-44.9463 & AGN \\
\hline 
82.6541&-37.5029 & AGN \\
\hline 
86.6358&-47.4166 & AGN\\
\hline 
98.5535&-51.3420 & AGN\\
\hline
144.6752 &20.0739 & AGN\\
\hline
179.1395 &-15.4257 & AGN\\
\hline
190.2065 &-28.8719 & AGN\\
\hline
202.8908 &-25.6041 & AGN\\
\hline
209.1916 &-18.7902 & AGN\\
\hline
210.4298 &-23.5555 & AGN\\
\hline
222.3923 &-18.7541 & AGN\\
\hline
135.8042 &-16.7185 & Low SNR/ Star?\\
\hline
\end{tabular}
\end{table}

\end{appendix}

\end{document}